%% file: main.tex
\documentclass[a4paper,fleqn]{cas-dc}

\usepackage[numbers]{natbib}
\usepackage{mdframed}
\usepackage{xcolor}
\usepackage{tikz}
\usepackage{multirow}
\usepackage{slashbox}
\usepackage[normalem]{ulem}

\usepackage{etoolbox}

\newcommand{\researchquestion}[2]{
    \vspace{10pt}
    \begin{mdframed}[backgroundcolor=yellow!20,
                    innermargin=1cm,
                    outermargin=1cm,
                    font=\small,]
        \textbf{#1} #2
    \end{mdframed}
    \vspace{10pt}
    }

\def\tsc#1{\csdef{#1}{\textsc{\lowercase{#1}}\xspace}}
\tsc{WGM}
\tsc{QE}
\tsc{EP}
\tsc{PMS}
\tsc{BEC}
\tsc{DE}

\begin{document}
\let\WriteBookmarks\relax
\def\floatpagepagefraction{1}
\def\textpagefraction{.001}

\shorttitle{Explaining the Contributing Factors for Vulnerability Detection in Machine Learning}

\title [mode = title]{Explaining the Contributing Factors for Vulnerability Detection in Machine Learning}                      



\shortauthors{Esma Mouine et~al.}

\author[1]{Esma Mouine}
\ead{e_mouine@encs.concordia.ca}
\author[1]{Yan Liu}[orcid=0000-0002-6747-8151]
\ead{yan.liu@concordia.ca}

\author[2]{Lu Xiao}
\ead{lxiao6@stevens.edu}

\author[3]{Rick Kazman}
\ead{kazman@hawaii.edu}

\author[2]{Xiao Wang}
\ead{xwang97@stevens.edu}

\address[1]{Concordia University,
1455 De Maisonneuve Blvd. W.
Montreal, Qc, Canada, 
H3G 1M8}

\address[2]{Stevens Institute of Technology,
Castle Point Terrace,
Hoboken, NJ 07030,
United States}

\address[3]{University of Hawaii, 
2500 Campus Rd, 
Honolulu, HI 96822, 
United States}












\begin{abstract}
There is an increasing trend to mine vulnerabilities from software repositories and use machine learning techniques to automatically detect software vulnerabilities.  A fundamental but unresolved research question is: how do different factors in the mining and learning process impact the accuracy of identifying vulnerabilities in software projects of varying characteristics? Substantial research has been dedicated in this area, including source code static analysis, software repository mining, and NLP-based machine learning. However, practitioners lack experience regarding the key factors for building a baseline model of the state-of-the-art. In addition, there lacks of experience regarding the transferability of the vulnerability signatures from project to project. This study investigates how the combination of different vulnerability features and three representative machine learning models impact the accuracy of vulnerability detection in 17 real-world projects. We examine two types of vulnerability representations: 1) code features extracted through  NLP with varying tokenization strategies and three different embedding techniques (bag-of-words, word2vec, and fastText)  and 2) a set of eight architectural metrics that capture the abstract design of the software systems. The three machine learning algorithms include a random forest model, a support vector machines model, and a residual neural network model. Overall, 95\% of the learning metrics  (precision, recall, and f1 score, etc.) are above 0.77 in the experiments out of 10 hypothesis tests and 408 experiments. Further analysis shows a recommended baseline model with signatures extracted through bag-of-words embedding, combined with the random forest, consistently increases the detection accuracy by about 4\% compared to other combinations in all 17 projects. Furthermore, we observe the limitation of transferring vulnerability signatures across domains based on our experiments.   

\end{abstract}




\maketitle

\section{INTRODUCTION}\label{sec:intro}
\input{Sections/Intro}

\section{RELATED WORK}\label{sec:related}
\input{Sections/Related}

\section{RESEARCH METHODOLOGY}\label{sec:rqs}
\input{Sections/RQ_methodology}
\section{EXPERIMENTS AND RESULTS}\label{sec:results}
\input{Sections/Experiments_Results}
\section{CROSS VALIDATION (RQ5)}
\input{Sections/Cross_validation}\label{sec:cross}
\section{DISCUSSION}\label{sec:dis}
\input{Sections/Discussions}
\section{THREATS TO VALIDITY}\label{sec:threat}
\input{Sections/Threats}

\input{Sections/Conclusion}

\bibliographystyle{model1-num-names}

\bibliography{mybibfile,arch}


\end{document}

%% file: Sections/Intro.tex
The National Institute of Standards and Technology (NIST) defines security vulnerability as a \textit{weakness in an information system, system security procedures, internal controls, or implementation that could be exploited or triggered by a threat source}~\cite{VulNIST}.  Software vulnerability management is the practice of identifying, classifying, remediating, and mitigating vulnerabilities. Early detection of vulnerable code reduces the risks of run-time errors, faults, threats, and the collapse of a system. As software scales expand,  vulnerability detection with sufficient accuracy and efficiency remains a challenge from both research ~\cite{dam2017automatic, li2018vuldeepecker, Ghaffarian:2017:SVA:3135069.3092566, shin2010evaluating} and industrial perspectives~\cite{russell2018automated, zimmermann2010searching}. The goal is to learn from representations of vulnerable features and to automate the discovery of vulnerabilities in source code.  

In industrial practice, security flaws are regularly reported to the Common Vulnerabilities and Exposures (CVE) database ~\cite{cve}. This database is used to collect and share publicly disclosed information about security vulnerabilities.  Likewise, Common Weakness Enumeration (CWE) is a community-developed list of common software and hardware security weaknesses~\cite{cwe}. The Open Web Application Security Project (OWASP) Benchmark is a Java test suite that contains thousands of exploitable test cases where each one maps to a specific CWE. NIST's Software Assurance Reference Dataset (SARD)~\cite{sard} provides a set of known security flaws for researchers and software security assurance developers. Within SARD, a set of test suites exist including the Juliet Tests for Java and C++~\cite{juliet_java:2017}, mobile apps and Web apps~\cite{sard-apps}. These sources of information are used to search for known vulnerabilities to identify potential exploits as part of a forensics process.

In software engineering, static code analysis helps to identify bugs or flaws in software.  Code analysis techniques are embedded in security scanners and raise alerts when vulnerabilities are detected~\cite{secubat2006,burp-suite,acunetix, netsparker,ffinder,checkmarx, RATS}. The identified vulnerabilities are confirmed by security engineers.  One technique of static code analysis is pattern matching~\cite{ffinder,checkmarx, RATS} that searches based on a set of rules. These rules, usually defined by security experts, enumerate known vulnerabilities. One limitation of scanners based on static analysis is the high false-positive rate~\cite{high-fpr}. For example, one case study~\cite{high-fpr} performed using a static analysis tool on Java source files showed that 45.7\% of discovered vulnerabilities were false positives. 

To improve the precision and recall of detecting vulnerabilities, research has been conducted to build a feature engineering methodology. Dam et al.~\cite{dam2017automatic}  used a Long Short Term Memory (LSTM) model to capture relationships between code elements. Likewise, Russel et al.~\cite{russell2018automated} developed a fast and scalable vulnerability detection tool for C and C++ based on deep feature representation learning that interprets source code.   Hovsepyan et al.~\cite{hovsepyan2012software} analyzed Java source code using bag-of-words and support vector machines to classify vulnerabilities.

Recent research has focused on machine learning models to mine feature representations from software repositories~\cite{Perl:2015:VFP:2810103.2813604}. In addition to machine learning, other methods based on natural language processing (NLP) have emerged. To extract features, these techniques treat the source code as a form of text. Software repositories contain code that forms the \textit{corpus} upon which feature representations can be learned. The concept of a corpus,  originating in linguistics, is a collection of text in one or more languages.  In NLP, the corpus is used to train learning models. For example, in the classic Word2Vec~\cite{mikolov2013efficient} model,  a corpus is used to produce the embedding of tokens that forms the relations of these tokens to each other in a multi-dimensional space. Zhou and Sharma~\cite{Zhou:2017:AIS:3106237.3117771} use commit messages and bug reports from repositories to identify software flaws.

The representation of software code as tokens does not contain the code dependencies and structural complexity. Software architecture metrics measure the complexity of software  entities~\cite{feng2016towards,sachitano2004security,sohr2010idea,almorsy2013automated,schwanke2013measuring,mo2016ICSE}. For example, \textit{Fan-In} and \textit{Fan-Out} of source files and classes are shown to impact the propagation of software quality issues through the inter-dependencies among software entities~\cite{schwanke2013measuring}. Due to the intrinsic connections between software architecture and security, prior studies have investigated how software architecture impacts the security of a system~\cite{feng2016towards,sachitano2004security,sohr2010idea,almorsy2013automated}. In this paper, we observe whether software architecture metrics are dominating contributor to vulnerability classification combined with token embedding without structural representation.

Finally, it remains unclear how transferable the identified signatures from one set of projects are able to detect the vulnerability of other projects. To test the vulnerability signatures, we need a baseline model to output the vulnerability classification. Such a model helps to establish a base to investigate techniques on feature representation, learning models, factors such as code structure, and complexity in learning vulnerability patterns. A baseline model is commonly used as in the artificial intelligence community \cite{baseline,google-baseline}. A baseline model serves as a reference point to compare the performances of other models that are usually more complex. A baseline model relies on the understanding of the key factors contributing to the discovery of vulnerability signatures through a combination of techniques and machine learning models.

In this paper, we assume tokens in a software repository form the corpus used to learn vulnerability patterns.These tokens are further embedded as numerical features for learning vulnerability classification. The ultimate goal is to develop a learning method that takes input as code embeddings from project repositories so that the learning is transferable to other projects. Hence, the core research question is: 

 \researchquestion{}{What are the contributing factors in learning processes that impact the accuracy of identifying vulnerabilities across software projects?}

Our paper concentrates on four aspects of the learning process: (A1) the \textbf{tokenization}  of the source code, (A2) \textbf{the generation of embeddings}, (A3) \textbf{architectural metrics} and (A4) \textbf{machine learning models}. For  \textbf{tokenization} we experiment with two tokenization approaches---with and without symbols and comments. For \textbf{embeddings}, we investigate three types of embedding methods (namely bag-of-words~\cite{bow}; word2vec ~\cite{mikolov2013efficient}, and fastText~\cite{DBLP:journals/corr/JoulinGBM16}).  For \textbf{architectural metrics}, we explore  eight file-based metrics to augment vulnerability representations. These metrics are introduced in detail in Section~\ref{sec:architectural_metrics}. They measure how source files are connected to each other in a system. We aim to examine whether adding architectural metrics helps to improve detection accuracy. For~\textbf{machine learning models}, we build three machine learning algorithms including a weak learner-based model (random forest~\cite{598994}), a kernel vector-based model (support vector machines~\cite{Cortes:1995:SN:218919.218929}), and a neural network model (residual neural network~\cite{he2015deep}). 

We evaluate the combined effects of the above four aspects on the accuracy of vulnerability classification over 17 java projects. These combinations led to a total of 408 experiments to collect all results and evaluate the ten hypotheses derived. The accuracy of the learning results is measured by six metrics, including precision, recall, f-measure, false-positive rate, the area under the precision-recall curve, and the area under the receiver operating characteristic curve.

Our results indicate that \( 95\% \) of the learning metrics are above \(0.77\) over all experiments. Further analysis shows that feature representations derived from the source code including all tokens, using bag-of-words embedding, combined with the random forest model, consistently increases the detection accuracy by about \(4\%\) compared to other combinations in all 17 projects listed in Table~\ref{tab:projects-datasets}. This learning model is further evaluated in a transfer learning context on 5 out of 15 Android projects in Table~\ref{tab:projects-datasets} with both precision and recall above 0.8. Comparing the features of token embeddings and the architecture metrics, token embeddings contribute more to the vulnerability classification. We observe that the combination of token pre-processing, conventional NLP embedding, and random forest model is sufficient for building the baseline of learning vulnerability with comparable performance to deep learning models including the structure of ResNet or LSTM\cite{dam2017automatic}. Such a baseline provides a reference point to quantify the minimal expected performance that the new vulnerability learning model should achieve.

%% file: Sections/Related.tex
\subsection{Machine Learning and Natural Language Processing for Detecting Vulnerabilities}

Our research aims to identify the factors contributing to the learning of software vulnerabilities from source code repositories. 
Several approaches have been developed aiming to improve the detection of vulnerabilities (e.g., \cite{yamaguchi2011vulnerability, dam2017automatic, hovsepyan2012software, 7424372, MILOSEVIC2017266}). One example is applying pattern recognition techniques to detect malware \cite{MalwareImages}. This technique \cite{MalwareImages} consists of visualizing malware binary gray-scale images and classifying these images according to observations that show that malware from the same families appears to be very similar in layout and texture. 

To build a vulnerability prediction model, the selection of features is essential. The most frequent features used in previous works are software metrics \cite{radjenovic2013software, 544352, Nagappan:2006:MMP:1134285.1134349} and developers activity \cite{shin2010evaluating}. 
Basili et al.\cite{544352} used source code metrics to classify the C++ code into binary code vulnerabilities back to 1996. Nagappan et al. \cite{Nagappan:2006:MMP:1134285.1134349} used complexity metrics like module metrics that consist of the number of classes, functions and variables in the module M, in addition to per-function and per-class metrics. They used those metrics with some Microsoft systems to identify faulty components. Perl et al. \cite{Perl:2015:VFP:2810103.2813604} considered metrics from developer activities by analyzing if commits were related to a vulnerability or not.  The methodology of this work \cite{Perl:2015:VFP:2810103.2813604} consists of combining machine learning using a support vector machine (SVM) classifier with code metrics gathered from repository metadata.

Recent work treats code as a form of text and uses natural language processing based methods for code analysis. Zhou and Sharma \cite{Zhou:2017:AIS:3106237.3117771} used commit messages and bug reports from repositories to identify software flaws using NLP techniques such as word2vec to create the embeddings used as features and machine learning classifiers. Hovsepyan et al.\cite{hovsepyan2012software} analyzed Java source code from Android applications using a bag-of-words representation and SVM for vulnerability prediction.

Pang et al.~\cite{7424372} further include n-grams in the feature vectors and used SVM for classification.  Jackson and Bennett \cite{jackson2018locating} using the Python Natural Language Toolkit \\ (NLTK)  to develop a machine learning agent that uses NLP techniques to convert the code to a matrix and identify a specific flaw---SQL injection---in Java byte code using decision trees and random forests for classification.

Other works focus more on using deep learning techniques such as Russel et al. \cite{russell2018automated} attempts to identify vulnerabilities using C and C++ source code at the function level based on deep feature representation learning that directly interprets lexed source code and also Dam et al.\cite{dam2017automatic} present an approach based on deep learning using an LSTM model, to automatically learn both semantic and syntactic features of code.

Apart from the work of Hovsepyan et al. \cite{hovsepyan2012software} most of these approaches focus on the feature engineering part, like Russel et al. \cite{russell2018automated} that uses a convolutional neural network to build the feature vectors.

A recent survey \cite{9108283} summarizes the techniques, the datasets and results obtained from vulnerability detection research that uses machine learning. According to their categories, our work falls in the text-based category, since we use a convolutional neural network (Resnet). 

Our approach focuses on detecting vulnerabilities in source code using machine learning and natural language processing techniques. However, we use general NLP-based techniques (bag-of-word, word2vec, fastText, and tokenizing code) associated with different machine learning models to identify the key factors contributing to the learning of the software flaws from code.

\subsection{Software Architecture and Security}  \label{sec:related-arch} 

Software architecture is the high-level abstract of a software system. Poor software architectural decisions are responsible for various software quality problems. Numerous previous research has underscored the impact of software architecture on security. 

Software architecture is the most important determinant to systematically achieve quality attributes in a software system, including software security~\cite{brosig2011automated}. Software security is, for many systems, the most important quality attribute driving the design.

Due to the intrinsic connections between software architecture and security, prior studies have investigated how software architecture impacts the security of a system~\cite{feng2016towards,sachitano2004security,sohr2010idea,almorsy2013automated}. However, little work has investigated how to leverage software architecture characteristics and metrics in machine learning processes to discover vulnerabilities. Researchers in software architecture have developed some measures to capture the complexity of software architecture entities~\cite{feng2016towards,sachitano2004security,sohr2010idea,almorsy2013automated,schwanke2013measuring,mo2016ICSE}. For example, \textit{Fan-In} and \textit{Fan-Out} of source files and classes are shown to impact the propagation of software quality issues through the inter-dependencies among software entities~\cite{schwanke2013measuring}. What remains unclear is whether and how different architecture metrics can be used as vulnerability representations for machine learning models to detect software vulnerabilities.

Previous research mostly focused on security assessment and evaluation from an architectural perspective. For example, Feng et al. found that software vulnerabilities are highly correlated with flawed architectural connections among source files~\cite{feng2016towards}.  Sohr and Berger found that software architecture analysis helps to concentrate on security-critical software modules and detect certain security flaws at the architectural level, such as the circumvention of APIs or incomplete enforcement of access control~\cite{Architecture-Centric-Security}. Brian and Issarny showed how software architecture benefits security by encapsulating security-related requirements at design-time~\cite{Security-benefits}. Antonino et al. ~\cite{architecture-level-security-evaluation} evaluated the security of existing service-oriented systems on the architectural level. Their method is based on recovering security-relevant facts about the system and interactive security analysis at the structural level.  Alkussayer and Allen~\cite{Security-risk-analysis} proposed a security risk evaluation approach by leveraging the architectural model of a system, assuming that components propagate their security risks to higher-level components in the architecture model. Alkussayer and Allen~\cite{scenario-based-security} assessed the level of security supported by a given architecture and qualitatively compared multiple architectures with respect to their security support.

Despite the high recognition of an architecture's impact on security, the is little focus on using architectural metrics as vulnerability signatures for machine learning models~\cite{Mellado2010ACO,Jain2011ARO}. Alshammari et al. ~\cite{Security-Metrics} is one of the few studies that investigated security metrics based on the composition, coupling, extensibility, inheritance, and design size of an object-oriented project. However, these metrics have not been compared with other vulnerability signatures, such as code features extracted using NLP. In addition, these metrics tightly tie into object-oriented concepts and may not be easy to transfer to other programming paradigms.
Motivated by the work of Feng et al., our study focuses on \emph{eight} architectural metrics that capture how software elements, i.e. source files, are interdependent on each other~\cite{feng2016towards}. And these metrics are generally applicable to software projects of different characteristics, such as the programming language used. In addition, although they are measured at the file level in this work, it is easy to roll up and down to the component level or method level following the same rationale to detect vulnerabilities at different granularities in future studies. Most importantly, to the best of our knowledge, we are the first to compare architectural metrics with code features extracted through NLP as vulnerability representations.

%% file: Sections/RQ_methodology.tex
The research method considers the learning task as a classification problem to the vulnerability signature. We consider four relevant aspects to the learning process, including (A1) tokenization: With regards to how tokens are extracted from software, we consider questions such as if code comments as tokens impact detection results; (A2) embedding: Tokens are transformed into numerical values. The effects of different embedding techniques are investigated; (A3) architectural metrics: We focus on architectural metrics that measure the complexity of the inter-dependencies among fine-grained software architecture elements at the file level. We consider eight architecture metrics which will be detailed later; lastly, (A4) machine learning algorithms for classification. 

We consider software as a corpus to develop the feature representation through token encoding. 
The tokens are the terms from the software coded separated according to the spaces and special characters. For the corpus, we use software code from open repositories. Then the encodings are embedded in machine learning models for vulnerability detection on datasets such as OWASP benchmark, Juliet test suite for Java, and Android Study. The architectural metrics are used as additional feature representations, along with code-based representations. Based on the above rationale, we propose to answer the following research questions:

\researchquestion{RQ1}{How does the filtering of tokens affect  source code vulnerability detection?}

When using NLP techniques to extract features, an essential preprocessing step is the tokenization of the source code. This step involves separating the code into tokens before creating the embeddings. Generally,   special symbols (including { } , . ; : [ ] ) ( + - = | \& ! ? * ˆ \textbackslash < > @ " ' ~ \# \%) should be filtered out from  the source code before separating it into tokens. Another question is: do the comments contain meaningful features and affect the features representations? To answer this question, we compare the performance of vulnerability detection with and without comments. 

\researchquestion{RQ2}{Does a specific embedding technique perform better across software projects?}

Embeddings are the process that maps each token to one vector, and the vector values are learned using a class of techniques such as bag-of-words~\cite{bow}, word2vec~\cite{mikolov2013efficient} and fastText~\cite{bojanowski2016enriching}. This research question evaluates whether a particular embedding technique constantly improves the performance of vulnerability detection across all 17 software projects.

\researchquestion{RQ3}{Can architectural metrics that measure the structural complexity of software improve vulnerability detection?}
We answer this question in two ways. First, we compare the learning performance separately using the NLP-based token embedding and using the architectural metric representation, respectively. Next, we merge these representations into the learning process to observe if the combination improves vulnerability detection compared to using either of them alone.

\researchquestion{RQ4}{Which machine learning model performs better across different projects?
}

We compare three kinds of machine learning models, namely, decision tree based (Random Forests), kernel-based (Support Vector Machines), and deep neural networks (Residual Neural Networks). Each model is combined with the feature representation extracted through different techniques of tokenization, embedding techniques, and architectural metrics. The goal is to discover whether a particular machine learning model performs best in terms of vulnerability detection in different settings and across software projects.

\researchquestion{RQ5}{How transferable is the learning in predicting vulnerabilities of projects in cross-validation?}
For this last research question, we aim to evaluate the transferability of the learned features in vulnerability prediction. The learning model is fine-tuned by training projects in cross-validation. 
We define different sets of experiment where we train our model with a project and predict the vulnerabilities on other projects. 

\subsection{The Vulnerability Detection Process}    
\begin{figure}[h]
    \centering
        \resizebox{\columnwidth}{!}{
        \input{Sections/fig/Learning_task.tikz}
        }
        \caption{Learning software  vulnerability as a classification task}
        \label{fig:learning-process}
    \end{figure}
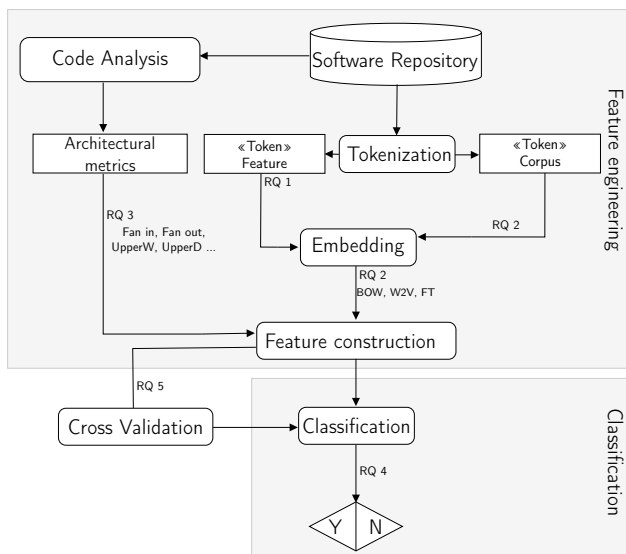

The process of vulnerable code detection, as shown in Figure~\ref{fig:learning-process},  contains a software repository, which provides the corpus for developing a vocabulary. Any project (even without vulnerable code labels) can be used for this purpose. Such a vocabulary is used to build the embedding of software tokens. The embedding is pretrained using word2Vec and fastText with the corpus. The tokens are then converted to numerical representations by running the embedding.  In addition, architecture metrics can be extracted from projects with tags. Next, architectural metrics and embeddings of code tokens are the features input to a supervised classification model.  We consider the vulnerability detection under two sources of vulnerability code labels: (1) the labels are from code within the same domain as the target software for vulnerability detection (Tables \ref{tab:results-all-tokens}, \ref{tab:results-no-comms} and \ref{tab:results-arch}); and (2) the models are trained with software code in one domain with vulnerability labels and used to classify software code in a different domain. For example, a model is learnt with the dataset from the Juliet dataset, then used to predict the vulnerability of source code in Android projects (Tables \ref{tab:cross-project-validation-vs-lstm} and \ref{tab:cross-domain}).

    

\subsection{Tokenization}
Tokenization is a common pre-processing step in natural language processing to transform the raw input text into a format that is more easily processed. The raw code contains 1) special symbols include punctuation characters (such as, . , : ; ? ) ( ] [ ~ ' " \} \{), 2) mathematical and logical operators (such as, + - / = * \& ! \% | < >); and 3) others (such as \# \textbackslash @ ˆ), in NLP special characters add no value to text-understanding and can induce noise in algorithms. In addition to other meta text that usually appears as code comments. These comments are any text that starts with two forward slashes (//) and any text between /* and */. 
To determine if those special characters and comments are important in the vulnerability prediction in source code, we use regular expressions to remove the code comments and special symbols.

\subsection{Token Embeddings}
Token embeddings are learned numerical representations for text where tokens or words that have similar meanings are approximated by the same value. In the domain of NLP, a corpus is a collection of texts. All tokens or words in the multiple corpora form a high-dimensional space. A learning model on this space calibrates the positions of each word or token according to its relations with all other tokens. Finally, each token has a numerical vector representation called an embedding. 

In this work, the corpus is formed with software projects selected from the Github repository.  We trained three models as follows to create the numerical vector representation of the source code tokens: 
\subsubsection{Bag-of-words (BOW)}
Bag-of-words is a representation of the text~\cite{bow}. It represents the text as a vector where each element is an index of a token from the vocabulary.  Each token is associated with its frequency in the text. Hence, the resulting vector has the same length as the number of unique tokens. The BOW vectors are limited to be the size of the text that is used for the training.
   
\subsubsection{Word2vec}
Word2vec is a method to create word embeddings that have been around since 2003~\cite{mikolov2013efficient}. The algorithm uses a neural network associated with a large corpus of text. Word2vec can use skip-gram or CBOW to learn the representations of tokens. Given a context, CBOW is the same as BOW, but instead of using sparse vectors (a vector with a lot of 0) to represent words, it uses dense vectors. CBOW predicts the probability of a target word. Skip-gram aims to predict the context of a word given its surrounding words. We use Skip-gram since we want to maintain the context of the token. The model takes a target term and creates a numerical vector from the surrounding terms.

\subsubsection{FastText}
FastText is a library for learning of word embeddings and text classification created by Facebook's AI research lab~\cite{DBLP:journals/corr/JoulinGBM16}. Akin to word2vec, fastText supports CBOW and skip-gram. Instead of feeding individual tokens into the neural network, fastText exploits the subterms information, which means each token is represented as a bag of characters in addition to the token itself. This allows the handling of unknown tokens, which aids cases where we want to take into account the internal structure of the words and handle unseen words.

Word2vec \& fastText use the same parameters. We use a dimensionality of 300 for the feature vectors and a window size of 5 and words with a total frequency lower than two are ignored. To obtain the source code embeddings of the files, we average the token vectors of all the terms of the file using tf-idf\footnote{Term Frequency - Inverse Document Frequency} weighting. These embeddings are calculated by multiplying each vector by the tf-idf weight of the corresponding term before calculating the average.
   
The resulting vector is the length of the vocabulary size. Our feature extractor uses the Python scikit-learn \cite{scikit-learn} library to generate the bag-of-words vector and the Gensim \cite{gensim} library for word2vec and fastText models. To train these models, we use source code from large repositories to learn the similarities between the source code tokens. The vocabulary is created from the source code of three large projects:  (1) the IntelliJ community project\cite{jetbrains_2019}, (2) the Android repository \cite{android-dev} and (3) the Android framework project. These repositories contain more than 70,000 Java files.

\subsection{Architectural Metrics}
    \label{sec:architectural_metrics}
    Software architecture refers to software elements, their relationships, and the properties of both \cite{bass2012software}. As discussed in Section~\ref{sec:related-arch}, prior research has revealed the significant impact of architecture design decisions on software security. In particular, the study in ~\cite{feng2016towards} reported that complicated architectural connections among source files in a project contribute positively to the propagation of software vulnerability issues. Hence we are motivated to investigate whether metrics that measure the complexity of architectural connections at the file level contribute positively to the detection of software vulnerabilities using machine learning models. 
    
    We model a set of architectural connections as a graph, namely $G = \{F, D\}$, where $F$ is the set of source files in the system, and $D$ is the set of structural dependencies among the source files. The graph $G$ of a software system can be reverse-engineered using existing tools, such as Scitool Understand~\footnote{\url{https://scitools.com/}}. 
    
    For each source file $f \in F$, we capture \emph{eight} metrics to measure the file's connections with the rest of the system $G$. We assume these metrics are feature representations to learn more vulnerabilities. These eight metrics are from three different but related aspects of software architecture:
    
     First, we measure \emph{Fan-in} and \emph{Fan-out} of a file $f$, which counts the number of direct dependencies with $f$, and are commonly used for various analysis:
    
    \begin{enumerate}
        \item Fan-in: The number of source files in $G$ that directly depends on $f$.
        \item Fan-out: The number of source files in $G$ that $f$ directly depends on.
    \end{enumerate}
    
    Next, we measure the position of $f$ in the entire dependency hierarchy of $G$. Cai et al. proposed an algorithm to cluster source files into hierarchical dependency layers based on their structural dependencies in $G$~\cite{cai2013leveraging}. The key features of the layers are: 1) source files in the same layers form independent modules, and 2) source files in a lower layer structurally depend on the upper layer, but not vice-versa. This layered structure is called the Architectural Design Rule Hierarchy (ArchDRH). The rationale is that source files in a higher layer structurally impact source files in the lower layers. Therefore, the higher the layer of $f$, the more influential it is for the rest of the system.
    
    \begin{enumerate}
    \setcounter{enumi}{2}
        \item Design Rule Hierarchy Layer: the layer number of $f$ in the ArchDRH clustering. 
    \end{enumerate}
    
    Finally, we measure the complexity of the transitive connections to each $f$ in $G$. For any $f \in F$, we define the $Butterfly\_Space_{f} = \{f, UpperWing, LowerWing\}$, where $f$ is the center of the space. $UpperWing$ is the set of source files that directly and transitively depend on $f$. Similarly, $LowerWing$ is the set of source files that $f$ directly and transitively depends on. For any $f \in G$, we calculate five metrics based on the $Butterfly\_Space$ notions:
    
    \begin{enumerate}
    \setcounter{enumi}{3}
        \item Space Size: the total number of source files in \\ $Butterfly\_Space_{f}$. This measures the total number of source files that $f$ is connected to directly and transitively. The higher this value, the more significant is $f$ connected to the rest of the system.
        
        \item Upper Width: the width of the $UpperWing$. This measures the maximal number of branches that depend on $f$.  
        
        \item Upper Depth: the length of the longest path in the $UpperWing$. This measures the most far-reaching transitive dependency on $f$. 
        
        \item Lower Width: the width of the $LowerWing$. This measures the maximal number of branches that $f$ depends on.  
        
        \item Lower Depth: the length of the longest path in the $LowerWing$. This measures the most far-reaching transitive dependency from $f$.
        
    \end{enumerate}
    
In this study, we investigate whether and to what extent these metrics contribute to the learning of software vulnerabilities.

\subsection{Machine Learning Models}
We perform a classification task to predict if a file is vulnerable or not. Our objective is to observe the effects of machine learning models. Since a machine learning model is part of the decision process of the classification task, we consider the model's transparency to the classification decision. A random forest model has one form of transparency as the feature importance to the classification performance. A kernel-based Support Vector Machine is useful for data with irregular distribution or unknown distribution. The residual neural network (ResNet) model has been used to examine explainability methods\cite{gilpin2019explaining}. Since the LSTM model was studied in the literature, we compare three kinds of machine learning models, decision tree-based Random Forests, kernel-based SVMs and deep neural networks as ResNet.
    \label{sec:pre-models}
        \subsubsection{Random Forest}
            The Random forest (RF) is an ensemble learning method for supervised classification~\cite{598994}. This model is constructed from multiple random decision trees. Those decision trees vote on how to classify a given instance of input data, and the random forest bootstraps those votes to prevent overfitting. 
        
        \subsubsection{Support Vector Machines}
            Support Vector Machines (SVM) uses a kernel function to perform both linear and non-linear classifications~\cite{Cortes:1995:SN:218919.218929}. The SVM algorithm creates a hyper-plane in a high-dimensional space that can separate the instances in the training set according to their class labels. 
            SVM is one of the widely used machine learning algorithms for sentiment analysis in NLP.

        \subsubsection{Residual Neural Network}
\begin{figure*}[pos=htbp,align=\centering]
     \centering
    \includegraphics[width=\textwidth]{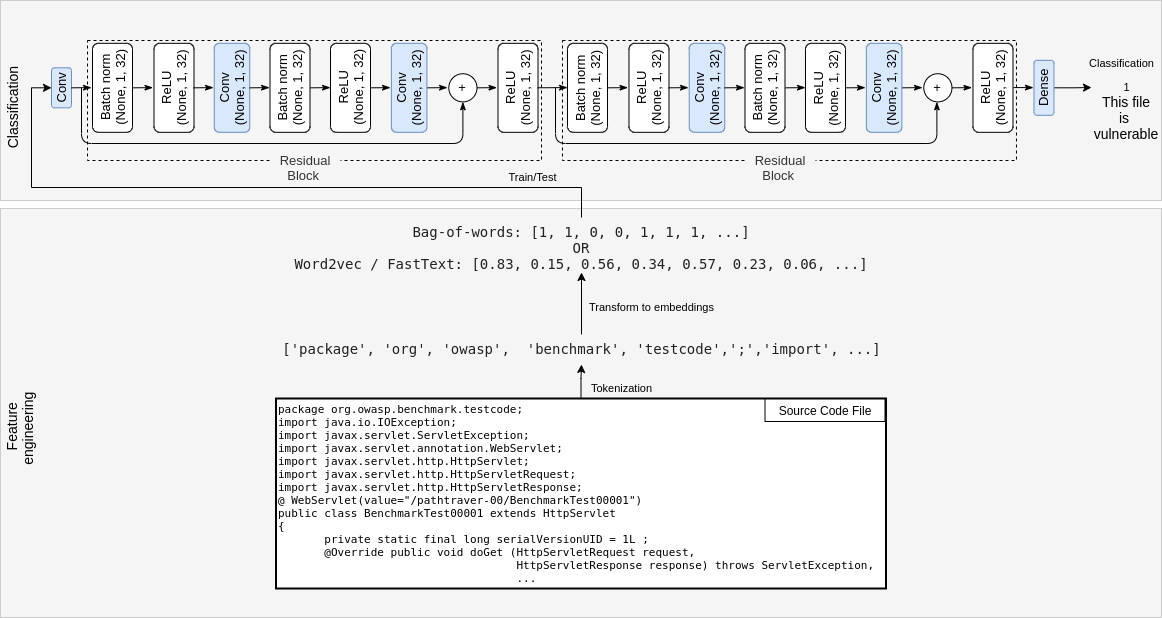}
    \caption{The data flow of the feature engineering and learning. The feature engineering is consistent for all the classification models. The modeling part illustrates the structure of the revised ResNet model that consists of one convolutional layer, one dense layer and 7 ResNet blocks. Each ResNet block is composed of 16 layers.}
    \label{fig:resNet_block}
\end{figure*}

            Residual Neural Network (ResNet) is a deep neuronal network model with residual blocks carrying linear data between neural layers. In our case, we construct the structure of a ResNet model composed of one convolutional layer, one dense layer and 7 ResNet blocks. Each ResNet block is composed of $16$ layers. The detailed ResNet structure is depicted in Figure ~\ref{fig:resNet_block}. We apply a residual block with the structure as follows:
            \begin{equation}
                \mathbf{x}_{l+1}= h\left(\mathbf{x}_{l}\right)+\mathcal{F}\left(\hat{f}\left(\mathbf{x}_{l}\right), \mathcal{W}_{l}\right)
            \end{equation}
            
            Where $\mathbf{x}$ is the input to the residual block and $\mathbf{l}$  indicates the $\mathbf{l-th}$ residual block. $\mathbf{\hat{f}}$ is the activation function which we use ReLU here. $\mathbf{F}$ is the residual function that contains two $1 \times 3$ convolutional layers. $\mathbf{W}$ stands for the corresponding parameters. We define the short cut $\mathbf{h}$ as one $1 \times 1$ convolutional layer if the dimension of $\mathbf{x}_{l}$ and $\mathbf{x}_{l+1}$ doesn't match, otherwise $\textit{h}$ will be:
            \begin{equation}
                h\left(\mathbf{x}_{l}\right)=\mathbf{x}_{l}
            \end{equation}
          
%% file: Sections/fig/Learning_task.tikz
\tikzset{every picture/.style={line width=0.75pt}} 

\begin{tikzpicture}[x=0.75pt,y=0.75pt,yscale=-1,xscale=1]

\draw  [color={rgb, 255:red, 204; green, 204; blue, 204 }  ,draw opacity=1 ][fill={rgb, 255:red, 245; green, 245; blue, 245 }  ,fill opacity=1 ] (344.5,416) -- (751.5,416) -- (751.5,608) -- (344.5,608) -- cycle ;
\draw  [color={rgb, 255:red, 204; green, 204; blue, 204 }  ,draw opacity=1 ][fill={rgb, 255:red, 245; green, 245; blue, 245 }  ,fill opacity=1 ] (80.5,17) -- (751.5,17) -- (751.5,405) -- (80.5,405) -- cycle ;
\draw  [color={rgb, 255:red, 0; green, 0; blue, 0 }  ,draw opacity=1 ] (661.75,193.83) -- (661.63,263) -- (582.29,262.89) ;
\draw [color={rgb, 255:red, 0; green, 0; blue, 0 }  ,draw opacity=1 ]   (582.29,262.89) -- (527,262.89) ;
\draw [shift={(524,262.89)}, rotate = 360] [fill={rgb, 255:red, 0; green, 0; blue, 0 }  ,fill opacity=1 ][line width=0.08]  [draw opacity=0] (8.93,-4.29) -- (0,0) -- (8.93,4.29) -- cycle    ;

\draw [color={rgb, 255:red, 0; green, 0; blue, 0 }  ,draw opacity=1 ]   (563.5,174.15) -- (587.5,174.15) ;
\draw [shift={(590.5,174.15)}, rotate = 180] [fill={rgb, 255:red, 0; green, 0; blue, 0 }  ,fill opacity=1 ][line width=0.08]  [draw opacity=0] (8.93,-4.29) -- (0,0) -- (8.93,4.29) -- cycle    ;
\draw  [fill={rgb, 255:red, 255; green, 255; blue, 255 }  ,fill opacity=1 ] (593.01,41.88) -- (593.01,88.92) .. controls (593.01,94.49) and (551.59,99) .. (500.51,99) .. controls (449.42,99) and (408,94.49) .. (408,88.92) -- (408,41.88) .. controls (408,36.31) and (449.42,31.8) .. (500.51,31.8) .. controls (551.59,31.8) and (593.01,36.31) .. (593.01,41.88) .. controls (593.01,47.44) and (551.59,51.96) .. (500.51,51.96) .. controls (449.42,51.96) and (408,47.44) .. (408,41.88) ;
\draw [color={rgb, 255:red, 0; green, 0; blue, 0 }  ,draw opacity=1 ]   (438.79,173.57) -- (426.26,173.18) ;
\draw [shift={(423.26,173.09)}, rotate = 361.78] [fill={rgb, 255:red, 0; green, 0; blue, 0 }  ,fill opacity=1 ][line width=0.08]  [draw opacity=0] (8.93,-4.29) -- (0,0) -- (8.93,4.29) -- cycle    ;
\draw [color={rgb, 255:red, 0; green, 0; blue, 0 }  ,draw opacity=1 ]   (183.88,96) -- (183.88,146.74) ;
\draw [shift={(183.88,149.74)}, rotate = 270] [fill={rgb, 255:red, 0; green, 0; blue, 0 }  ,fill opacity=1 ][line width=0.08]  [draw opacity=0] (8.93,-4.29) -- (0,0) -- (8.93,4.29) -- cycle    ;
\draw [color={rgb, 255:red, 0; green, 0; blue, 0 }  ,draw opacity=1 ]   (501.83,98) -- (501.33,150.26) ;
\draw [shift={(501.3,153.26)}, rotate = 270.55] [fill={rgb, 255:red, 0; green, 0; blue, 0 }  ,fill opacity=1 ][line width=0.08]  [draw opacity=0] (8.93,-4.29) -- (0,0) -- (8.93,4.29) -- cycle    ;
\draw [color={rgb, 255:red, 0; green, 0; blue, 0 }  ,draw opacity=1 ]   (457.3,294.54) -- (457.3,349.63) ;
\draw [shift={(457.3,352.63)}, rotate = 270] [fill={rgb, 255:red, 0; green, 0; blue, 0 }  ,fill opacity=1 ][line width=0.08]  [draw opacity=0] (8.93,-4.29) -- (0,0) -- (8.93,4.29) -- cycle    ;
\draw [color={rgb, 255:red, 0; green, 0; blue, 0 }  ,draw opacity=1 ]   (457.3,487.78) -- (457.3,547.61) ;
\draw [shift={(457.3,550.61)}, rotate = 270] [fill={rgb, 255:red, 0; green, 0; blue, 0 }  ,fill opacity=1 ][line width=0.08]  [draw opacity=0] (8.93,-4.29) -- (0,0) -- (8.93,4.29) -- cycle    ;
\draw  [fill={rgb, 255:red, 255; green, 255; blue, 255 }  ,fill opacity=1 ] (457.3,549.61) -- (507.53,576.11) -- (457.3,602.6) -- (407.07,576.11) -- cycle ;
\draw    (457.3,549.61) -- (457.3,602.6) ;
\draw  [fill={rgb, 255:red, 255; green, 255; blue, 255 }  ,fill opacity=1 ] (591.24,151.77) -- (721.84,151.77) -- (721.84,193.78) -- (591.24,193.78) -- cycle ;
\draw  [color={rgb, 255:red, 0; green, 0; blue, 0 }  ,draw opacity=1 ] (290,368) -- (183.88,368) -- (183.88,194.83) ;
\draw [color={rgb, 255:red, 0; green, 0; blue, 0 }  ,draw opacity=1 ]   (290,368) -- (345,367.05) ;
\draw [shift={(348,367)}, rotate = 539.01] [fill={rgb, 255:red, 0; green, 0; blue, 0 }  ,fill opacity=1 ][line width=0.08]  [draw opacity=0] (8.93,-4.29) -- (0,0) -- (8.93,4.29) -- cycle    ;
\draw  [fill={rgb, 255:red, 255; green, 255; blue, 255 }  ,fill opacity=1 ] (108,151.77) -- (274.78,151.77) -- (274.78,193.78) -- (108,193.78) -- cycle ;
\draw  [color={rgb, 255:red, 0; green, 0; blue, 0 }  ,draw opacity=1 ] (216.5,448) -- (215.83,383.68) -- (349.14,382.29) ;
\draw  [fill={rgb, 255:red, 255; green, 255; blue, 255 }  ,fill opacity=1 ] (293.18,152.08) -- (423.78,152.08) -- (423.78,194.1) -- (293.18,194.1) -- cycle ;
\draw  [color={rgb, 255:red, 0; green, 0; blue, 0 }  ,draw opacity=1 ] (354.58,194.53) -- (354.15,273.9) -- (372.21,274) ;
\draw [color={rgb, 255:red, 0; green, 0; blue, 0 }  ,draw opacity=1 ]   (372.21,274) -- (391.93,274) ;
\draw [shift={(394.93,274)}, rotate = 180] [fill={rgb, 255:red, 0; green, 0; blue, 0 }  ,fill opacity=1 ][line width=0.08]  [draw opacity=0] (8.93,-4.29) -- (0,0) -- (8.93,4.29) -- cycle    ;
\draw  [fill={rgb, 255:red, 255; green, 255; blue, 255 }  ,fill opacity=1 ] (439.3,161.26) .. controls (439.3,156.84) and (442.88,153.26) .. (447.3,153.26) -- (556.5,153.26) .. controls (560.92,153.26) and (564.5,156.84) .. (564.5,161.26) -- (564.5,185.26) .. controls (564.5,189.67) and (560.92,193.26) .. (556.5,193.26) -- (447.3,193.26) .. controls (442.88,193.26) and (439.3,189.67) .. (439.3,185.26) -- cycle ;
\draw  [fill={rgb, 255:red, 255; green, 255; blue, 255 }  ,fill opacity=1 ] (397.3,262.26) .. controls (397.3,257.84) and (400.88,254.26) .. (405.3,254.26) -- (514.5,254.26) .. controls (518.92,254.26) and (522.5,257.84) .. (522.5,262.26) -- (522.5,286.26) .. controls (522.5,290.67) and (518.92,294.26) .. (514.5,294.26) -- (405.3,294.26) .. controls (400.88,294.26) and (397.3,290.67) .. (397.3,286.26) -- cycle ;
\draw  [fill={rgb, 255:red, 255; green, 255; blue, 255 }  ,fill opacity=1 ] (350,363.26) .. controls (350,358.84) and (353.58,355.26) .. (358,355.26) -- (557.5,355.26) .. controls (561.92,355.26) and (565.5,358.84) .. (565.5,363.26) -- (565.5,387.26) .. controls (565.5,391.67) and (561.92,395.26) .. (557.5,395.26) -- (358,395.26) .. controls (353.58,395.26) and (350,391.67) .. (350,387.26) -- cycle ;
\draw  [fill={rgb, 255:red, 255; green, 255; blue, 255 }  ,fill opacity=1 ] (395.3,455.26) .. controls (395.3,450.84) and (398.88,447.26) .. (403.3,447.26) -- (512.5,447.26) .. controls (516.92,447.26) and (520.5,450.84) .. (520.5,455.26) -- (520.5,479.26) .. controls (520.5,483.67) and (516.92,487.26) .. (512.5,487.26) -- (403.3,487.26) .. controls (398.88,487.26) and (395.3,483.67) .. (395.3,479.26) -- cycle ;
\draw [color={rgb, 255:red, 0; green, 0; blue, 0 }  ,draw opacity=1 ]   (457.3,394.78) -- (457.3,443) ;
\draw [shift={(457.3,446)}, rotate = 270] [fill={rgb, 255:red, 0; green, 0; blue, 0 }  ,fill opacity=1 ][line width=0.08]  [draw opacity=0] (8.93,-4.29) -- (0,0) -- (8.93,4.29) -- cycle    ;
\draw [color={rgb, 255:red, 0; green, 0; blue, 0 }  ,draw opacity=1 ]   (406.88,67) -- (291,67) ;
\draw [shift={(288,67)}, rotate = 360] [fill={rgb, 255:red, 0; green, 0; blue, 0 }  ,fill opacity=1 ][line width=0.08]  [draw opacity=0] (8.93,-4.29) -- (0,0) -- (8.93,4.29) -- cycle    ;
\draw  [fill={rgb, 255:red, 255; green, 255; blue, 255 }  ,fill opacity=1 ] (94,56.8) .. controls (94,51.53) and (98.28,47.26) .. (103.55,47.26) -- (278.95,47.26) .. controls (284.22,47.26) and (288.5,51.53) .. (288.5,56.8) -- (288.5,85.45) .. controls (288.5,90.72) and (284.22,95) .. (278.95,95) -- (103.55,95) .. controls (98.28,95) and (94,90.72) .. (94,85.45) -- cycle ;
\draw [color={rgb, 255:red, 0; green, 0; blue, 0 }  ,draw opacity=1 ][fill={rgb, 255:red, 0; green, 0; blue, 0 }  ,fill opacity=1 ][line width=0.75]    (302.5,469) -- (391.5,469) ;
\draw [shift={(394.5,469)}, rotate = 180] [fill={rgb, 255:red, 0; green, 0; blue, 0 }  ,fill opacity=1 ][line width=0.08]  [draw opacity=0] (8.93,-4.29) -- (0,0) -- (8.93,4.29) -- cycle    ;
\draw   (135.5,456.26) .. controls (135.5,451.84) and (139.08,448.26) .. (143.5,448.26) -- (294.5,448.26) .. controls (298.92,448.26) and (302.5,451.84) .. (302.5,456.26) -- (302.5,480.26) .. controls (302.5,484.67) and (298.92,488.26) .. (294.5,488.26) -- (143.5,488.26) .. controls (139.08,488.26) and (135.5,484.67) .. (135.5,480.26) -- cycle ;

\draw (499.69,72.74) node  [font=\LARGE] [align=center] {Software Repository};
\draw (504.48,173.09) node  [font=\LARGE] [align=center] {Tokenization};
\draw (459.48,273.78) node  [font=\LARGE] [align=center] {Embedding};
\draw (451.48,375.46) node  [font=\LARGE] [align=center] {Feature construction};
\draw (457.3,576.11) node  [font=\LARGE] [align=center] {Y \ \ \ N};
\draw (656.54,172.78) node  [font=\large] [align=center] {<<Token>> \\ Corpus};
\draw (191.39,172.78) node  [font=\Large] [align=center] {Architectural \\  metrics};
\draw (356.58,197.53) node [anchor=north west][inner sep=0.75pt]  [font=\normalsize,color={rgb, 255:red, 0; green, 0; blue, 0 }  ,opacity=1 ] [align=center] {RQ 1};
\draw (459.78,509.45) node [anchor=north west][inner sep=0.75pt]  [font=\normalsize,color={rgb, 255:red, 0; green, 0; blue, 0 }  ,opacity=1 ] [align=center] {RQ 4};
\draw (186,233) node [anchor=north west][inner sep=0.75pt]  [font=\normalsize,color={rgb, 255:red, 0; green, 0; blue, 0 }  ,opacity=1 ] [align=center] {RQ 3};
\draw (459.3,297.54) node [anchor=north west][inner sep=0.75pt]  [font=\normalsize,color={rgb, 255:red, 0; green, 0; blue, 0 }  ,opacity=1 ] [align=center] {RQ 2};
\draw (457.9,467.26) node  [font=\LARGE] [align=center] {Classification};
\draw (191.25,71.13) node  [font=\LARGE] [align=center] {Code Analysis};
\draw (500.69,323.74) node  [font=\small,color={rgb, 255:red, 0; green, 0; blue, 0 }  ,opacity=1 ] [align=center] {BOW, W2V, FT};
\draw (249.69,266.74) node  [font=\normalsize,color={rgb, 255:red, 0; green, 0; blue, 0 }  ,opacity=1 ] [align=center] {Fan in, Fan out, \\ UpperW, UpperD ...};
\draw (218.5,418) node [anchor=north west][inner sep=0.75pt]  [font=\normalsize,color={rgb, 255:red, 0; green, 0; blue, 0 }  ,opacity=1 ] [align=center] {RQ 5};
\draw (219,468.26) node  [font=\LARGE] [align=center] {Cross Validation};
\draw (603.3,243.54) node [anchor=north west][inner sep=0.75pt]  [font=\normalsize,color={rgb, 255:red, 0; green, 0; blue, 0 }  ,opacity=1 ] [align=center] {RQ 2};
\draw (358.48,173.09) node  [font=\large] [align=center] {<<Token>> \\ Feature};
\draw (748,100) node [anchor=north west][inner sep=0.75pt]  [font=\LARGE,rotate=-90] [align=center] {Feature engineering \ };
\draw (748,446) node [anchor=north west][inner sep=0.75pt]  [font=\LARGE,rotate=-90] [align=center] {Classification};

\end{tikzpicture}

%% file: Sections/Experiments_Results.tex
Our evaluations are based on the tasks of 1) defining hypotheses to answer each research question; 2) preparing appropriate datasets; 3) defining metrics to evaluate the learning effects, and 4) running experiments and collecting results to test the hypotheses. The hypotheses test if tokenization techniques, embedding techniques, architectural metrics and machine learning models have significant effects on the ability to learn software vulnerabilities. 

\subsection{Datasets}
We prepare three datasets with vulnerabilities labelled, including the OWASP Benchmark project~\cite{owasp}, the Juliet test suite for Java~\cite{juliet_java:2017} and 15 Android applications from the previous Android study~\cite{androidStudy}. OWASP and Juliet have the vulnerability types available online. Android study follows the labels published in the paper~\cite{androidStudy}. 
    
\subsubsection{OWASP Benchmark Project}
The OWASP Benchmark is a free test suite designed to evaluate automated software vulnerability detection tools. It contains 2740 test cases with 1415 vulnerable files (52\%) and 1325 non-vulnerable files (48\%). Table~\ref{tab:owasp-vul} enumerates the different types of vulnerabilities found in the OWASP project. 
    
\begin{table}
    \centering
    \caption{OWASP vulnerability Types}
    \label{tab:owasp-vul}
    \begin{tabular}{|l|c|c|}
        \hline
        \textbf{Vulnerability Area} & \textbf{CWE} & \textbf{\# of files} \\ \hline \hline
        Command Injection          & 78  & 251         \\ \hline
        Weak Cryptography          & 327 & 246         \\ \hline
        Weak Hashing               & 328 & 236         \\ \hline
        LDAP Injection             & 90  & 59          \\ \hline
        Path Traversal             & 22  & 268         \\ \hline
        Secure Cookie Flag         & 614 & 67          \\ \hline
        SQL Injection              & 89  & 504         \\ \hline
        Trust Boundary Violation   & 501 & 126         \\ \hline
        Weak Randomness            & 330 & 493         \\ \hline
        XPath Injection            & 643 & 35          \\ \hline
        XSS (Cross-Site Scripting) & 79  & 455         \\ \hline
    \end{tabular}
\end{table}

\subsubsection{Test Suite for Java}
This test suite contains 217 vulnerable files (58\%) and 297 non-vulnerable files (42\%). There are 112 different vulnerabilities and errors such as buffer overflow, OS injection, hard-coded password, absolute path traversal, NULL pointer dereference, uncaught exception, deadlock, missing releases of resource and others listed in Table~\ref{tab:juliet-vul}.  
\begin{table}[h!]
    \centering
    \caption{Juliet Test Suite Vulnerability Types}
    \label{tab:juliet-vul}
    \resizebox{\columnwidth}{!}{
    \begin{tabular}{|l|c|c|}
        \hline
       \textbf{Vulnerability Area}                                             & \textbf{CWE} & \textbf{\# of files} \\ \hline \hline
        Integer Overflow or Wraparound                                                & 190 & 115 \\ \hline
        Integer Underflow                                                             & 191 & 92  \\ \hline
        Improper Validation of Array Index                                            & 129 & 72  \\ \hline
        SQL Injection                                                                 & 89 & 60   \\ \hline
        Divide By Zero                                                                & 369 & 50  \\ \hline
        Uncontrolled Memory Allocation                                                & 789 & 42  \\ \hline
        Uncontrolled Resource Consumption                                             & 400 & 39  \\ \hline
        HTTP Response Splitting                                                       & 113 & 36  \\ \hline
        Numeric Truncation Error                                                      & 197 & 33  \\ \hline
        Basic Cross-site scripting                                                    & 80 & 18   \\ \hline
        Use of Externally-Controlled Format String                                    & 134 & 18  \\ \hline
        XPath Injection                                                               & 643 & 12  \\ \hline
        Assignment to Variable without Use                                            & 563 & 12  \\ \hline
        Unchecked Input for Loop Condition                                            & 606 & 12  \\ \hline
        OS Command Injection                                                          & 78  & 12  \\ \hline
        Relative Path Traversal                                                       & 23  & 12  \\ \hline
        Unsafe Reflection                                                             & 470 & 12  \\ \hline
        LDAP Injection                                                                & 90  & 12  \\ \hline
        Absolute Path Traversal                                                       & 36  & 12  \\ \hline
        Configuration Setting                                                         & 15  & 12  \\ \hline
        \multicolumn{2}{|l|}{Others}                                                        & 67  \\  \hline 
    \end{tabular}
    }
\end{table}

As shown in Table~\ref{tab:owasp-vul} and  Table~\ref{tab:juliet-vul}, the common vulnerabilities between the three datasets are the SQL injection (CWE 89) and the command injection (CWE 78). The vulnerabilities in common between OWASP and Juliet are command injection (CWE 78), LDAP injection (CWE 90), SQL injection and XPATH injection (CWE 643). And the vulnerability type that we can find in all three projects is Cross-site scripting (CWE 79 \& 80).

\subsubsection{Android Study}
The \emph{Android Study} is a public dataset that contains 20 different Java applications that cover a variety of domains. This dataset is used in the work of  Scandariato \textit{et al}~\cite{6860243}. According to~\cite{6860243},  the source code was scanned using the Fortify Source Code Analyzer, a security scanning tool to mark the vulnerable files. In total, the Android Study contains $2321$ vulnerable files such as cross-site scripting, SQL injection, header manipulation, privacy violation and command injection. The label is binary that is vulnerable or not, without the exact type of vulnerability for each file. We collect the information of the application names, the versions and the paths of the file with its vulnerable label.  Using these references, we develop scripts to retrieve 15 projects for our evaluation. Table~\ref{tab:projects-datasets} shows the 17 applications we use in this project and the vulnerability rate of the labelled source code for each. Since Fortify itself may produce errors in the vulnerability scanning, the quality of labelling is not fully evaluated. This is a potential threat to validity.

\begin{table}[ht]
    \centering
    \caption{Dataset Vulnerability Statistics}
    \label{tab:projects-datasets}
    \resizebox{\columnwidth}{!}{
    \begin{tabular}{|ll|c|c|c|}
        \hline
        \multicolumn{2}{|c|}{Projects} & Vulnerability rate & Number of files & \# of tokens \\
        \hline \hline
        1 & QuickSearchBox & 23\% & 654 & 4301 \\
        2 & FBReader & 30\% & 3450 & 6589 \\
        3 & Contacts & 31\% & 787 & 13438 \\
        4 & Browser & 37\% & 433 & 9561 \\
        5 & Mms & 37\% & 865 & 7965 \\
        6 & Camera & 38\% & 475 & 7851 \\
        7 & KeePassDroid & 39\% & 1580 & 2872 \\
        8 & Calendar & 44\% & 307 & 8003 \\
        9 & ConnectBot & 46\% & 104 & 4109 \\
        10 & Crosswords & 46\% & 842 & 4223 \\
        11 & K9 & 47\% & 2660 & 13175 \\
        12 & Deskclock & 47\% & 127 & 2163 \\
        13 & Coolreader & 49\% & 423 & 5424 \\
        14 & OWASP & 52\% & 2740 & 6154 \\
        15 & Email & 54\% & 840 & 15454 \\
        16 & Juliet & 58\% & 514 & 1268 \\
        17 & AnkiDroid & 59\% & 275 & 8408\\\hline
    \end{tabular}
    }
\end{table}

\begin{table}[h]
    \centering
    \caption{Common vulnerabilities in the three datasets}
    \resizebox{\columnwidth}{!}{
    \begin{tabular}{|l|c|c|c|c|}
        \hline
        Vulnerability Area & CWE & OWASP & Juliet & Android \\ \hline\hline
        Cross-Site Scripting & 79 & X & X & X \\ \hline
        SQL injection & 89 & X & X & X \\ \hline
        Command Injection & 78 &  &  & X \\ \hline
        XPath Injection & 643 & X & X &  \\ \hline
        OS Command Injection & 78 & X & X &  \\ \hline
        LDAP Injection & 90 & X & X &  \\ \hline
    \end{tabular}
    }
\end{table}
    
\subsection{Analysis of Tokens}

Each line of code is parsed to produce tokens including variables, preserved keywords, operators, symbols and separators.

\begin{figure}
    \centering
    \caption{OWASP token distribution. Most of the tokens have fewer than 20 occurrences.} \includegraphics[width=\columnwidth]{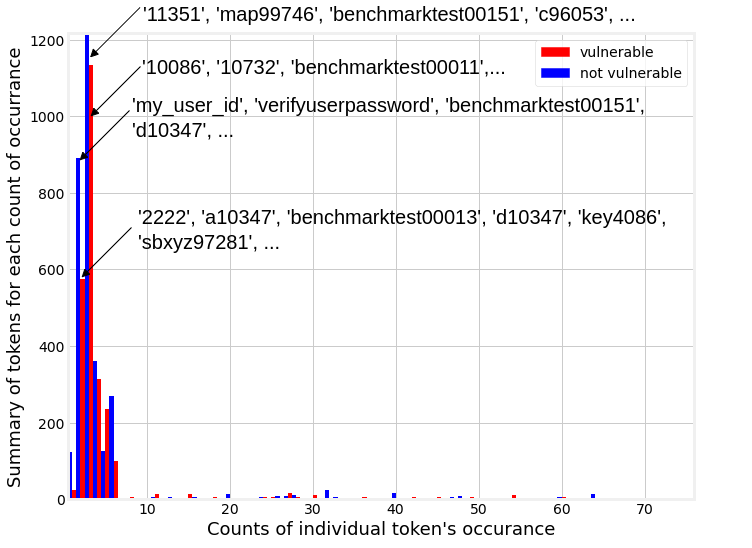}
    \label{fig:owasp-tokens-occ}
\end{figure}

\begin{figure}
        \centering
        \caption{Juliet token distribution. Most of the tokens are have fewer than 25 occurrences.} \includegraphics[width=\columnwidth]{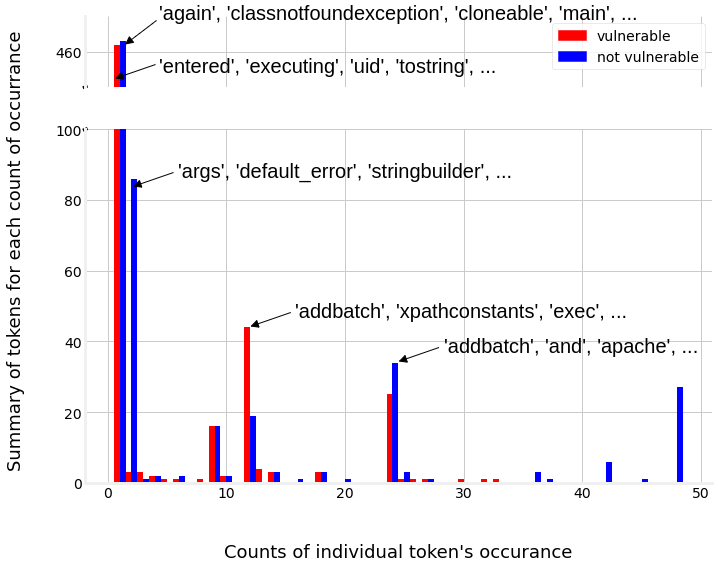}
        \label{fig:juliet-tokens-occ}
    \end{figure}

\begin{figure}
    \centering
    \caption{All Android projects token distribution. The majority of the tokens have fewer than 500 occurrences.}
    \includegraphics[width=\columnwidth]{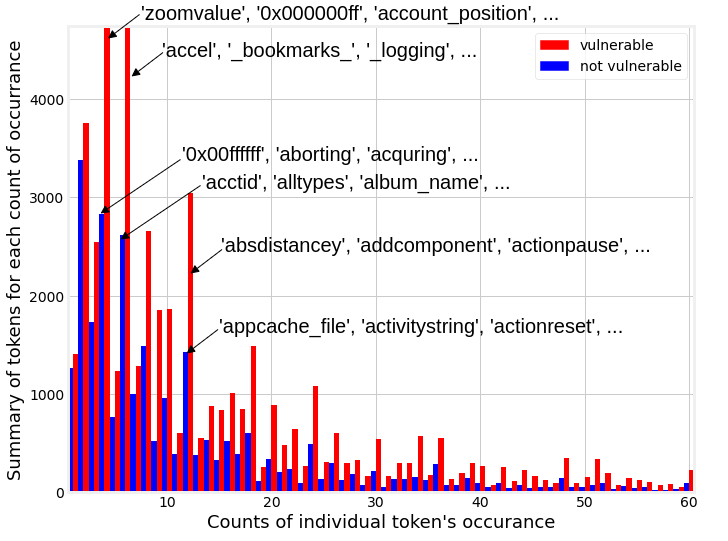}
    \label{fig:android-tokens-occ}
\end{figure}

First, we analyze the token statistics and observe if any significant characters of the tokens. For each dataset OWASP, Juilet, and Android project, we separate the tokens in vulnerable files from tokens in non-vulnerable files and plot the token frequency distribution in Figure~\ref{fig:owasp-tokens-occ}, Figure~\ref{fig:juliet-tokens-occ}, and Figure~\ref{fig:android-tokens-occ} respectively.  

For the OWASP project (shown in Figure~\ref{fig:owasp-tokens-occ}), tokens are mostly grouped in the counts of occurrence that are less than 20. Beyond 20 occurrences, the counts of tokens are significantly smaller. In Juliet source code (shown in Figure~\ref{fig:juliet-tokens-occ}), the distribution of the token frequency has more peaks than the OWASP token distribution.
In all the Android source code (shown in Figure~\ref{fig:android-tokens-occ}), the tokens are mostly grouped with occurrences less than 30. 

The charts show two facts: (1) the frequency distribution of each project varies. This could be a factor in downgrading the accuracy of cross domain learning; (2) the high-frequency tokens are neutral such as “main”, “string\_builder”. This indicates the feature representation is learnt mainly from tokens with less frequency. These two facts relate to the experimental results presented in section \ref{sec:results}. 
  
  \begin{table}[]
    \caption{Number of tokens in each dataset according to the vulnerability of the files and number of the common tokens in the vulnerable files and non vulnerable files.}
    \label{tab:tokens_number}
    \resizebox{\columnwidth}{!}{
    
    \begin{tabular}{|l|c|c|c|}
    \hline  
    \multirow{2}{*}{Dataset} & \# of tokens in & \# of tokens in & \# of tokens in \\ 
     & vulnerable files & non vulnerable files & common \\ \hline
    
    Owasp   & 2982                             & 3599                                 & 605                    \\ 
    Juliet  & 678                              & 764                                  & 460                    \\ 
    Android & 54196                            & 28698                                & 18339                  \\ \hline
    \end{tabular}
    }
\end{table}

\subsection{Evaluation Metrics}
\label{sec:eval-metrics}
For each experiment, we evaluate the performance of vulnerability detection using traditional Information Retrieval metrics. In the context of this study, True positives (TP) are the correct identification of source files with vulnerabilities. True negatives (TN) are the correct identification of source files without vulnerabilities. False positives (FP) are the incorrect identification of source files with vulnerabilities. False negatives (FN) are the incorrect identification of source files without vulnerabilities. Based on these metrics, we define the following metrics to measure the performance of vulnerability detection:

\begin{itemize}\setlength\itemsep{1em}
    \item The precision (P), which is the probability a file that is classified as vulnerable, is truly vulnerable.
    \begin{equation}
        P = \frac{TP}{(TP + FP)}
    \end{equation}
    
    \item The recall (R) which is the probability that a vulnerable sample of code is classified as vulnerable.
    \begin{equation}
        R = \frac{TP}{(TP + FN)}
    \end{equation}
    
    \item The f-measure (F1) is the harmonic average of precision and recall.
    
    \begin{equation}
        F1 = 2 * \frac{1}{\frac{1}{P} + \frac{1}{R}}
    \end{equation}
    
    \item The false positive rate (FPR) is the proportion of negative cases incorrectly identified as positive cases. 
    \begin{equation}
        FPR = \frac{FP}{(FP + TN)}
    \end{equation}
    
    \item The area under the precision-recall curve (PR AUC) summarizes the information in the precision-recall curve.
    
    \item The area under the receiver operating characteristic curve (ROC AUC) shows the capability of the model to distinguish between classes.
\end{itemize}

To aggregate the above values of metrics for comparing different hypotheses, we define an aggregation formula below. 
We first add all the metrics defined above in Eq(\ref{eq:sum}), $\forall s \in S \{P, R, F1, 1-FPR, ROC AUC, PR AUC\}$. We then normalize the value using Eq(\ref{eq:normalization}), where $N$ is the number of comparison cases. 

\begin{equation}
    \label{eq:sum}
    \mathbf{x}_{i}=\sum \mathbf{s}_{i}
\end{equation}

\begin{center}
    \begin{equation}
        \label{eq:normalization}
        \mathbf{z}_{i}=\frac{\mathbf{x}_{i} - \min_{\forall k \in [1,N]}(x_{k})}{\max_{\forall k \in [1,N]}(x_{k})-\min_{\forall k \in [1,N]}(x_{k})}
    \end{equation}
\end{center}

\subsection{Experiment Design and Hypotheses}
\label{sec:results}
To answer the research questions, we first design experiments for each project and then developed experiments for cross-project validation. In the first set of experiments, each project's source code is divided into training and testing partitions. The vulnerability detection models are trained per project and tested on the same project. 
Due to space limitations, we include the result tables in the appendix \ref{sec:appendix}. 
 
We use the $z_{i}$ to compute the p-value from a paired \textit{t test} for each comparison.  When $\alpha\geq0.5$ the hypothesis is accepted to be significantly different. Otherwise, the hypothesis is rejected. Table \ref{tab:p-value-results} shows the different p-values obtained for each hypothesis. A detailed analysis is presented in the following sections.

\begin{table*}[h]
    \centering
    \caption{p-value obtained from the \emph{t test} for the 10 hypotheses}
    \label{tab:p-value-results}
    \resizebox{\textwidth}{!}{
    \begin{tabular}{|l|c|c|c|c|c|c|c|c|c|c|}
    \cline{2-11}
    \multicolumn{1}{c|}{} & Tokenization & \multicolumn{3}{c|}{Embeddings} & \multicolumn{3}{c|}{Architectural metrics} & \multicolumn{3}{c|}{Models} \\ \hline
    Hypothesis & 1 & 2 & 3 & 4 & 5 & 6 & 7 & 8 & 9 & 10 \\ \hline
    p-value    & 0.01 & $1.02\mathrm{e}{-10}$ & $1.08\mathrm{e}{-06}$ & 0.69 & $1.4\mathrm{e}{-16}$ & $0.56$ & $7\mathrm{e}{-15}$ & $3.8\mathrm{e}{-06}$ & $3.8\mathrm{e}{-12}$ & $3\mathrm{e}{-05}$ \\ \hline
    Conclusion & Reject & Accept & Accept & Reject & Accept & Reject & Accept & Accept & Accept & Accept \\ \hline
    \end{tabular}
    }
\end{table*}

\subsection{Experiment Results for Tokenization (RQ1) } 
\textbf{Observation:} We observe whether tokenization with removing code comments and/or special symbols may improve or create noise for the detection. We run experiments with the tokens including the comments and symbols (Table~\ref{tab:results-all-tokens} in Appendix) compared to tokens without them (Table~~\ref{tab:results-no-comms}). Each table shows the learning scores of 153 experiments (= (3 models x 3 embeddings) per project x 17 projects). Each experiment produces six scores that compute the value of $z$. Totally 306 data points of  $z$ are used to compute p-value from the \textit{t test}. The p-value is compared to the significance level $\alpha$. 

\textbf{Hypothesis Analysis:} Hypothesis (1) is defined as follows:
\begin{enumerate}
    \item There is a statistically significant difference between the results obtained from using all tokens vs. using tokens without comments and symbols.
\end{enumerate}
According to Table~\ref{tab:p-value-results}, the p-value obtained for hypothesis (1) is less than the significance level 0.05. This hypothesis is then rejected, which means there is no statistically significant difference between the two tokenization strategies.

\textbf{Conclusion:} We conclude that comments and symbols do not affect the learning of software vulnerabilities from source code in our experiments.

\subsection{Experiment Results for Feature Extraction (RQ2)} 
\textbf{Observation:} Feature extraction techniques convert tokens into a vector of features. In this experiment, we observe the effects of three feature extraction techniques, including (1) bag-of-words, (2) word2vec embedding and (3) fastText. We run experiments with features obtained from bag-of-words (Table~\ref{tab:results-all-tokens} and Table~\ref{tab:results-no-comms}). For each embedding technique we have 102 experiments (= (3 models x 2 tokenization methods) per project x 17 projects). Each experiment produces six scores that compute the value of $z$. Totally 204 data points of $z$ are used to compute p-value from the \textit{t test}. The p-value is compared to the significance level $\alpha$. 

\textbf{Hypothesis Analysis:} To compare those three vector representation techniques, we consider the following hypotheses:
\begin{enumerate}
    \setcounter{enumi}{1}
    \item There is a statistically significant difference between the results obtained from using bag-of-words and \\ word2vec as embeddings.
    \item There is a statistically significant difference between the results obtained from using bag-of-words and fastText as embeddings.
     \item There is a statistically significant difference between the results obtained from using word2vec and fastText as embeddings.
\end{enumerate}

According to the Table~\ref{tab:p-value-results} the p-values obtained from \emph{t test} of hypothesis (2) and hypothesis (3) are accepted, but hypothesis (4) is rejected. This means there is a statistically significant difference between bag-of-words and word2vec and also between bag-of-words and fastText. However, there is no statistically significant difference between word2vec and fastText. Additionally, the results obtained from the classification show that, on average, the precision and recall of the experiments with bag-of-words are 6\% more than the performance of the other embedding methods. That indicates that bag-of-words is better than the other two models in the learning process of vulnerabilities in our experiments.

\textbf{Conclusion:} We choose bag-of-words as the best way to generate embeddings for the remainder of the experiments.

\subsection{Experiment Results using Architectural Metrics (RQ3)}

\textbf{Observation:} In addition to the NLP-based method, we use architectural metrics as structural features to learn vulnerability patterns from the software repositories.  We compare the features with code tokens only.
We run experiments using the architecture metrics only and compared them to using the architecture metrics with the bag-of-words features. The Table~\ref{tab:results-arch} in Appendix shows the learning score of 51 (= 3 models x 17 projects) experiments for each feature we use. Totally 102 data points of  $z$ are used to compute p-value from the \textit{t test}. The p-value is compared to the significance level $\alpha$. 

\textbf{Hypothesis Analysis:} The effects of using architecture metrics, extracted from the structures of the project, are explored via three hypotheses: 

\begin{enumerate}
    \setcounter{enumi}{4}
    \item There is a statistically significant difference between 1) the results obtained from using tokens, vs. 2) the results obtained from using the architectural metrics.
    \item There is a statistically significant difference between 1) the results obtained from only using tokens, vs. 2) the results obtained from using the combination of architectural metrics and tokens.
    \item There is a statistically significant difference between 1) the results obtained from only using the architectural metrics,  vs. 2) the results obtained from using the combination of tokens and architectural metrics. 
\end{enumerate}

As shown in the Table \ref{tab:p-value-results}, the p-values indicate hypotheses (5) and (7) are accepted, while hypothesis (6) is rejected. This means there is no significant difference when using tokens as input features with or without architectural metrics. Hypotheses (5) and (7) further indicate the tokens have a stronger influence on the learning performance than the architectural metrics.  

\textbf{Conclusion:} We choose to use tokens without architectural metrics for the remainder of the experiments. 

\subsection{Experiment Results on Classification models (RQ4) }
\textbf{Observation:} We aim to identify whether a certain machine learning model produces better vulnerability detection.
We run experiments with each of the three models to compare them (Table~\ref{tab:results-all-tokens} and Table~\ref{tab:results-no-comms} in Appendix). 
For each model, the table shows 102 experiments (= (2 tokenization methods x 3 embeddings) per project x 17 projects) Each experiment produces six scores that compute the value of $z$. Totally 204 data points of  $z$ are used to compute p-value from the \textit{t test}. The p-value is compared to the significance level $\alpha$.

\textbf{Hypothesis Analysis:} To compare the models, we define these three hypotheses:
\begin{enumerate}
    \setcounter{enumi}{7}
    \item There is a statistically significant difference between the performance of the random forest model and the SVM.
    \item There is a statistically significant difference between the performance of the random forest model and the ResNet.
    \item There is a statistically significant difference between the performance between the SVM and the ResNet.
\end{enumerate}
According to Table \ref{tab:p-value-results}, the p-values of the three hypotheses (8), (9), and (10) are less than the significance level, $\alpha$. All three hypotheses are accepted. Overall the random forest model performs better than the SVM and ResNet in most of the experiments with a precision and recall higher by an average of 8\%. 

\textbf{Conclusion:} We decide to use the random forest as the model to learn the patterns of vulnerabilities in the cross-project validation experiments. 

%% file: Sections/Cross_validation.tex
In this evaluation, we explore the answer to the question \emph{"How transferable is the learning method in predicting vulnerabilities in new projects?"}.
We define two sets of experiments to investigate this question.



\textbf{Train-One-Predict-Multiple} In this test, a learning model is trained with source code from a single project and then tested on other projects. We compare the learning performance with existing work~\cite{dam2017automatic}. The 15 projects used in this experiment overlap with those used in~\cite{dam2017automatic}. We use the same score to evaluate the learning performance as in ~\cite{dam2017automatic}:  we count the number of projects with the classification metrics of precision and recall with a certain threshold. Table \ref{tab:cross-project-validation-vs-lstm} reports comparison between our work and the learning with LSTM models.  With the threshold value of 0.7, our results are comparable to the results in ~\cite{dam2017automatic}. With a threshold of 0.8, our results degrade to the average value of 1.4 projects with both a precision and a recall equal to or greater than 80\%.

\begin{table}[h]
    \centering
    \caption{Training-One-Predicting-Multiple compared with the LSTM model in~\cite{dam2017automatic} with the threshold value settings.}
    \label{tab:cross-project-validation-vs-lstm}
    \resizebox{\columnwidth}{!}{
    \begin{tabular}{|ll|c|c|c|}
        \hline
        \multicolumn{2}{|c|}{\multirow{2}{*}{Projects}}  
        & \multicolumn{1}{c|}{Random Forest}
        & \multicolumn{1}{c|}{Random Forest}
        & \multicolumn{1}{c|}{LSTM\footnote{The experiments are performed using 18 Android project. In our case we only retrieved 15 of them that are still available}~\cite{dam2017automatic} }\\
              & &  (precision$>$70\%, & (precision$>$ 80\%, & (precision$>$80\%, \\
              & &  recall $>$ 70\%) & recall $>$ 80\%) & recall $>$ 80\%) \\ \hline
            1 &     Camera          & 7     & 4      & 6   \\ 
            2 &     FBReader        & 6     & 3      & 6   \\ 
            3 &     Mms             & 6     & 2      & 6   \\ 
            4 &     Contacts        & 6     & 2      & 2   \\ 
            5 &     KeePassDroid    & 6     & 2      & 4   \\ 
            6 &     ConnectBot      & 6     & 2      & 5   \\ 
            7 &     AnkiDroid       & 5     & 1      & 5   \\ 
            8 &     Email           & 5     & 0      & 4   \\ 
            9 &     Crosswords      & 4     & 1      & 1   \\ 
            10 &    Browser         & 4     & 1      & 1   \\ 
            11 &    Coolreader      & 4     & 1      & 6   \\ 
            12 &    Calendar        & 4     & 0      & 5   \\ 
            13 &    K9              & 3     & 2      & 8   \\ 
            14 &    DeskClock       & 0     & 0      & 1   \\ 
            15 &    QuickSearchBox  & 0     & 0      & 3   \\ \hline
        \end{tabular}
        }
\end{table}

\textbf{Train-Multiple-Predict-One} To further improve the learning performance, we conduct 15-fold cross-validation by choosing 14 projects from the same domain of the Android project for training. The remaining project is reserved for testing. Table \ref{tab:cross-project-validation} contains the  cross-validation results, ordered by precision and recall values. 5 out of the 15 experiments have both precision and recall values equal to or greater than 80\%; 10 out of 15 experiments have both precision and recall equal to or greater than 70\%. Referring to Table~\ref{tab:projects-datasets}, the 5 experiments with precision and recall below 70\% have the ratio of vulnerable files below 40\%. 

Referring to Table~\ref{tab:cross-project-validation}, cross-project validation improves the learning performance under the threshold of 80\% to 5 projects out of 15 projects. This approach of transfer learning, by combining the features from the Android project repository to tune the random forest model, achieves comparable learning performance to deep learning models ResNet and LSTM~\cite{dam2017automatic} (4.2 projects out of 15 projects). 


\begin{table}[h]
\centering
\caption{The cross project validation from 15 Android projects, with 5 projects having both precision and recall higher than 80\% (ConnectBot, Email, Coolreader, Crosswords, AnkiDroid) } 
\label{tab:cross-project-validation}
\resizebox{\columnwidth}{!}{
\begin{tabular}{|ll|cccccc|c|}
\hline
\multicolumn{2}{|l|}{Projects} & P & R & F1 & FPR & ROC AUC & {PR AUC} & {z} \\ \hline
1 & \textbf{ConnectBot} & \textbf{0.90} & \textbf{0.86} & 0.88 & 0.08 & 0.89 & 0.84 & 1.00 \\ \hline
2 & \textbf{Email} & \textbf{0.90} & \textbf{0.81} & 0.85 & 0.10 & 0.86 & 0.83 & 0.95 \\ \hline
3 & \textbf{Coolreader} & \textbf{0.88} & \textbf{0.82} & 0.85 & 0.11 & 0.86 & 0.81 & 0.93 \\ \hline
4 & \textbf{Crosswords} & \textbf{0.81} & \textbf{0.87} & 0.84 & 0.14 & 0.86 & 0.76 & 0.89 \\ \hline
5 & K9 & 0.94 & 0.60 & 0.74 & 0.05 & 0.78 & 0.78 & 0.81 \\ \hline
6 & \textbf{AnkiDroid} & \textbf{0.81} & \textbf{0.86} & 0.83 & 0.29 & 0.78 & 0.78 & 0.80 \\ \hline
7 & \textbf{Calendar} & \textbf{0.75} & \textbf{0.88} & 0.81 & 0.24 & 0.82 & 0.71 & 0.79 \\ \hline
8 & \textbf{Camera} & \textbf{0.74} & \textbf{0.87} & 0.80 & 0.32 & 0.77 & 0.71 & 0.72 \\ \hline
9 & \textbf{FBReader} & \textbf{0.73} & \textbf{0.71} & 0.72 & 0.11 & 0.80 & 0.61 & 0.68 \\ \hline
10 & Contacts & 0.69 & 0.92 & 0.79 & 0.39 & 0.77 & 0.67 & 0.67 \\ \hline
11 & KeePassDroid & 0.64 & 0.90 & 0.75 & 0.34 & 0.78 & 0.62 & 0.63 \\ \hline
12 & Deskclock & 0.64 & 0.88 & 0.74 & 0.33 & 0.77 & 0.61 & 0.61 \\ \hline
13 & Browser & 0.70 & 0.70 & 0.70 & 0.17 & 0.76 & 0.60 & 0.61 \\ \hline
14 & Mms & 0.68 & 0.70 & 0.70 & 0.20 & 0.76 & 0.59 & 0.59 \\ \hline
15 & QuickSearchBox & 0.45 & 0.93 & 0.60 & 0.46 & 0.74 & 0.44 & 0.38 \\ \hline
\end{tabular}
}
\end{table}

%% file: Sections/Discussions.tex
Our approach consists of comparing the different factors that contributed to the detection of vulnerabilities in source code.   The result tables contain the metrics for the different experiments that we have performed. In each experiment we used a Java project with a combination of the aspects explained in the previous sections of this paper. The vulnerability detection models are trained and tested with the same project. Each dataset is separated into a training set and a test set. 

\input{Sections/tab/result-table-wcomments}

\input{Sections/tab/result-table-nocomments}
\input{Sections/tab/results-table-arch}

Tables \ref{tab:results-all-tokens}, \ref{tab:results-no-comms} and \ref{tab:results-arch} show the results of the 408 experiments performed to evaluate the classification on three domains of the source code. Table \ref{tab:results-all-tokens} contains the results obtained from the experiments using the source code files of all tokens. Table \ref{tab:results-no-comms} are the results obtained from the experiments using the source code files without the comments and symbols. Table \ref{tab:results-arch} shows the results of using the architectural metrics as features compared to bag-of-words.

 These tables and the statistical test performed in section~\ref{sec:results} demonstrate 95\% of the learning metrics are above 0.77 after over 400 experiments. The tokenization choice, which consists of removing the comments or not, shows that the comments and symbols do not affect the learning of the vulnerabilities by the model. Using the architectural metrics as features, in this case, has no significant improvement on the learning of vulnerabilities. One reason is the complexity of the code and its dependencies are not captured by the tokens only.  As a result, a baseline emerges with feature representations extracted through bag-of-words embedding and using the random forest model. This baseline increases the accuracy by about 4\% compared to other combinations of examined factors.
 
We further conducted cross-validation to observe how transferable across domains the vulnerability signatures are. The training of a single project and predicting multiple projects method achieves an average of 4.4 projects with a precision and recall higher than 70\%. With the 15-fold cross-validation method of training multiple projects and predicting on one project,  the baseline model slightly outperforms the LSTM model with a proprietary embedding method~\cite{dam2017automatic}.







%% file: Sections/tab/result-table-wcomments.tex
\begin{table*}[h!]
    \centering
    \caption{Singular project vulnerability detection with tokenization with comments across embeddings and machine learning models.}
    \label{tab:results-all-tokens}
    \resizebox{\textwidth}{!}{
        \begin{tabular}{|ll|l|cccccc|c|cccccc|c|cccccc|c|}
            \cline{4-24}
            \multicolumn{1}{c}{} & \multicolumn{1}{c}{} & \multicolumn{1}{c}{} & \multicolumn{7}{|c|}{Bag-of-words} & \multicolumn{7}{c|}{Word2vec} & \multicolumn{7}{c|}{FastText}\\ \cline{1-24}
            \multicolumn{2}{|c|}{Project} & Classifier & P & R & F1 & FPR & ROC AUC & PR AUC & z & P & R & F1 & FPR & ROC AUC & PR AUC & z & P & R & F1 & FPR & ROC AUC & PR AUC & z \\ \cline{1-24}
            \multirow{3}{*}{1}  & \multirow{3}{*}{OWASP}     & RF & 1.00 & 1.00 & 1.00 & 0.21 & 1.00    & 1.00   & 0.95 & 0.68 & 0.76 & 0.72 & 0.21 & 0.70    & 0.63   & 0.60 & 1.00 & 1.00 & 1.00 & 0.21 & 1.00    & 1.00   & 0.95 \\ 
             & & ResNet & 0.99 & 0.92 & 0.95 & 0.10 & 0.96    & 0.95   & 0.92 & 0.95 & 0.93 & 0.94 & 0.04 & 0.94    & 0.92   & 0.92 & 0.76 & 0.96 & 0.84 & 0.04 & 0.85    & 0.87   & 0.82 \\
             & & SVM    & 0.99 & 0.99 & 0.99 & 0.24 & 0.99    & 0.99   & 0.93 & 0.91 & 0.93 & 0.92 & 0.19 & 0.93    & 0.89   & 0.86 & 0.82 & 0.90 & 0.86 & 0.08 & 0.93    & 0.92   & 0.85 \\ \hline
            \multirow{3}{*}{2}  & \multirow{3}{*}{Juliet}    & RF & 1.00 & 1.00 & 1.00 & 0.08 & 1.00    & 1.00   & 0.98 & 0.07 & 0.05 & 0.05 & 0.06 & 0.29    & 0.41   & 0.02 & 0.12 & 0.09 & 0.10 & 0.04 & 0.30    & 0.40   & 0.05 \\
             & & ResNet & 1.00 & 0.73 & 0.84 & 0.13 & 0.86    & 0.84   & 0.80 & 0.21 & 0.18 & 0.20 & 0.02 & 0.34    & 0.38   & 0.13 & 0.38 & 0.68 & 0.49 & 0.11 & 0.44    & 0.64   & 0.42 \\ 
             & & SVM    & 1.00 & 1.00 & 1.00 & 0.07 & 1.00    & 1.00   & 0.98 & 0.17 & 0.05 & 0.07 & 0.05 & 0.50    & 0.42   & 0.10 & 0.42 & 0.23 & 0.29 & 0.84 & 0.50    & 0.42   & 0.07 \\ \hline 
            \multirow{3}{*}{3}  & \multicolumn{1}{l|}{\multirow{3}{*}{AnkiDroid}}   & RF & 0.80 & 0.89 & 0.84 & 0.15 & 0.84    & 0.76   & 0.76 & 0.85 & 0.97 & 0.91 & 0.10 & 0.89    & 0.84   & 0.85 & 0.85 & 0.97 & 0.91 & 0.10 & 0.89    & 0.84   & 0.85 \\ 
             & \multicolumn{1}{l|}{} & ResNet & 0.79 & 0.85 & 0.82 & 0.21 & 0.82    & 0.75   & 0.72 & 0.88 & 0.50 & 0.64 & 0.08 & 0.71    & 0.71   & 0.62 & 0.80 & 0.13 & 0.23 & 0.06 & 0.55    & 0.57   & 0.35 \\ 
             & \multicolumn{1}{l|}{} & SVM    & 0.80 & 0.89 & 0.84 & 0.13 & 0.86    & 0.78   & 0.78 & 0.80 & 0.93 & 0.86 & 0.34 & 0.83    & 0.78   & 0.73 & 0.82 & 0.93 & 0.87 & 0.63 & 0.85    & 0.80   & 0.69 \\ \hline 
            \multirow{3}{*}{4}  & \multicolumn{1}{l|}{\multirow{3}{*}{Browser}} & RF & 0.97 & 0.93 & 0.95 & 0.08 & 1.00    & 1.00   & 0.95 & 0.97 & 0.94 & 0.95 & 0.06 & 0.96    & 0.93   & 0.92 & 0.89 & 1.00 & 0.94 & 0.06 & 0.97    & 0.92   & 0.92 \\ 
             & \multicolumn{1}{l|}{} & ResNet & 0.94 & 0.88 & 0.91 & 0.09 & 0.92    & 0.87   & 0.87 & 0.38 & 0.97 & 0.55 & 0.10 & 0.56    & 0.38   & 0.47 & 0.83 & 0.91 & 0.87 & 0.02 & 0.90    & 0.88   & 0.85 \\
             & \multicolumn{1}{l|}{} & SVM    & 0.91 & 0.94 & 0.93 & 0.10 & 0.94    & 0.88   & 0.88 & 0.82 & 0.90 & 0.86 & 0.17 & 0.90    & 0.78   & 0.79 & 0.88 & 0.72 & 0.79 & 0.46 & 0.90    & 0.85   & 0.69 \\ \hline
            \multirow{3}{*}{5}  & \multicolumn{1}{l|}{\multirow{3}{*}{Calendar}}       & RF & 0.87 & 0.87 & 0.89 & 0.00 & 0.92    & 0.82   & 0.85 & 0.89 & 0.86 & 0.88 & 0.00 & 0.88    & 0.84   & 0.85 & 1.00 & 0.86 & 0.93 & 0.00 & 0.92    & 0.95   & 0.92 \\ 
             & \multicolumn{1}{l|}{} & ResNet & 0.85 & 1.00 & 0.92 & 0.00 & 0.95    & 0.85   & 0.90 & 0.58 & 0.97 & 0.73 & 0.00 & 0.67    & 0.58   & 0.66 & 0.88 & 0.48 & 0.62 & 0.00 & 0.71    & 0.80   & 0.65 \\ 
             & \multicolumn{1}{l|}{} & SVM    & 0.88 & 0.95 & 0.91 & 0.00 & 0.94    & 0.85   & 0.89 & 0.86 & 0.86 & 0.86 & 0.00 & 0.87    & 0.81   & 0.83 & 0.84 & 0.72 & 0.78 & 0.00 & 0.88    & 0.91   & 0.80 \\ \hline 
            \multirow{3}{*}{6}  & \multicolumn{1}{l|}{\multirow{3}{*}{Camera}}  & RF & 0.94 & 0.91 & 0.93 & 0.08 & 0.94    & 0.89   & 0.89 & 0.89 & 0.83 & 0.86 & 0.08 & 0.89    & 0.80   & 0.81 & 0.82 & 0.75 & 0.78 & 0.11 & 0.95    & 0.84   & 0.77 \\ 
             & \multicolumn{1}{l|}{} & ResNet & 0.91 & 0.91 & 0.91 & 0.10 & 0.93    & 0.87   & 0.87 & 0.55 & 1.00 & 0.71 & 0.05 & 0.77    & 0.55   & 0.65 & 0.92 & 0.46 & 0.61 & 0.10 & 0.72    & 0.76   & 0.62 \\ 
             & \multicolumn{1}{l|}{} & SVM    & 0.91 & 0.94 & 0.93 & 0.04 & 0.94    & 0.88   & 0.90 & 0.82 & 0.77 & 0.79 & 0.06 & 0.80    & 0.67   & 0.72 & 0.72 & 0.75 & 0.73 & 0.37 & 0.90    & 0.82   & 0.66 \\ \hline 
            \multirow{3}{*}{7}  & \multicolumn{1}{l|}{\multirow{3}{*}{ConnectBot}}     & RF & 1.00 & 1.00 & 1.00 & 0.00 & 1.00    & 1.00   & 1.00 & 1.00 & 0.80 & 0.89 & 0.04 & 0.90    & 0.90   & 0.87 & 1.00 & 0.80 & 0.89 & 0.00 & 0.90    & 0.90   & 0.88 \\ 
             & \multicolumn{1}{l|}{} & ResNet & 1.00 & 1.00 & 1.00 & 0.00 & 1.00    & 0.90   & 0.98 & 1.00 & 0.07 & 0.13 & 0.12 & 0.53    & 0.55   & 0.33 & 1.00 & 0.80 & 0.89 & 0.12 & 0.90    & 0.90   & 0.85 \\ 
             & \multicolumn{1}{l|}{} & SVM    & 1.00 & 1.00 & 1.00 & 0.02 & 1.00    & 1.00   & 0.99 & 1.00 & 0.80 & 0.89 & 0.06 & 0.90    & 0.90   & 0.87 & 1.00 & 0.80 & 0.89 & 0.12 & 0.90    & 0.90   & 0.85 \\ \hline 
            \multirow{3}{*}{8}  & \multicolumn{1}{l|}{\multirow{3}{*}{Contacts}}       & RF & 0.90 & 0.96 & 0.93 & 0.11 & 0.96    & 0.88   & 0.89 & 0.83 & 0.86 & 0.85 & 0.06 & 0.89    & 0.76   & 0.80 & 0.89 & 1.00 & 0.94 & 0.06 & 0.99    & 0.97   & 0.94 \\ 
             & \multicolumn{1}{l|}{} & ResNet & 0.78 & 0.90 & 0.83 & 0.08 & 0.89    & 0.73   & 0.78 & 0.56 & 1.00 & 0.72 & 0.15 & 0.81    & 0.56   & 0.65 & 0.79 & 0.67 & 0.73 & 1.00 & 0.79    & 0.79   & 0.48 \\ 
             & \multicolumn{1}{l|}{} & SVM    & 0.90 & 0.96 & 0.93 & 0.21 & 0.96    & 0.88   & 0.86 & 0.80 & 0.78 & 0.79 & 0.36 & 0.87    & 0.77   & 0.69 & 0.85 & 0.80 & 0.82 & 1.00 & 0.93    & 0.82   & 0.58 \\ \hline 
            \multirow{3}{*}{9}  & \multicolumn{1}{l|}{\multirow{3}{*}{Coolreader}}     & RF & 1.00 & 0.98 & 0.99 & 0.09 & 1.00    & 1.00   & 0.97 & 1.00 & 1.00 & 1.00 & 0.03 & 1.00    & 1.00   & 0.99 & 1.00 & 1.00 & 1.00 & 0.05 & 1.00    & 1.00   & 0.99 \\ 
             & \multicolumn{1}{l|}{} & ResNet & 1.00 & 0.93 & 0.96 & 0.12 & 0.96    & 0.98   & 0.93 & 0.80 & 0.73 & 0.76 & 0.08 & 0.80    & 0.69   & 0.70 & 0.83 & 0.91 & 0.87 & 0.30 & 0.89    & 0.79   & 0.76 \\ 
             & \multicolumn{1}{l|}{} & SVM    & 0.97 & 0.97 & 0.97 & 0.03 & 0.96    & 0.93   & 0.95 & 1.00 & 1.00 & 1.00 & 0.11 & 1.00    & 1.00   & 0.97 & 1.00 & 1.00 & 1.00 & 0.39 & 1.00    & 1.00   & 0.91 \\ \hline 
            \multirow{3}{*}{10} & \multicolumn{1}{l|}{\multirow{3}{*}{Deskclock}}      & RF & 0.89 & 1.00 & 0.94 & 0.02 & 0.97    & 0.89   & 0.92 & 0.86 & 1.00 & 0.92 & 0.02 & 0.93    & 0.86   & 0.89 & 0.90 & 1.00 & 0.95 & 0.02 & 0.99    & 0.98   & 0.95 \\ 
             & \multicolumn{1}{l|}{} & ResNet & 0.88 & 0.88 & 0.88 & 0.03 & 0.91    & 0.80   & 0.84 & 0.46 & 1.00 & 0.63 & 0.05 & 0.50    & 0.46   & 0.53 & 0.35 & 1.00 & 0.51 & 0.01 & 0.50    & 0.67   & 0.54 \\ 
             & \multicolumn{1}{l|}{} & SVM    & 0.89 & 1.00 & 0.94 & 0.02 & 0.97    & 0.89   & 0.92 & 0.80 & 1.00 & 0.89 & 0.04 & 0.93    & 0.86   & 0.87 & 0.82 & 1.00 & 0.90 & 0.02 & 1.00    & 0.99   & 0.93 \\ \hline 
            \multirow{3}{*}{11} & \multicolumn{1}{l|}{\multirow{3}{*}{Email}}   & RF & 0.97 & 0.98 & 0.97 & 0.30 & 0.99    & 0.99   & 0.91 & 0.98 & 0.98 & 0.98 & 0.00 & 0.99    & 1.00   & 0.98 & 0.91 & 0.95 & 0.93 & 0.00 & 0.98    & 0.97   & 0.94 \\ 
             & \multicolumn{1}{l|}{} & ResNet & 0.96 & 0.90 & 0.93 & 0.23 & 0.93    & 0.96   & 0.87 & 0.74 & 0.88 & 0.81 & 0.33 & 0.75    & 0.84   & 0.69 & 0.76 & 0.91 & 0.83 & 0.56 & 0.80    & 0.86   & 0.67 \\ 
             & \multicolumn{1}{l|}{} & SVM    & 0.95 & 0.82 & 0.88 & 0.23 & 0.94    & 0.95   & 0.84 & 0.79 & 0.84 & 0.81 & 0.27 & 0.88    & 0.89   & 0.75 & 0.80 & 0.92 & 0.86 & 0.27 & 0.91    & 0.90   & 0.79 \\ \hline 
            \multirow{3}{*}{12} & \multicolumn{1}{l|}{\multirow{3}{*}{FBReader}}      & RF & 0.96 & 0.93 & 0.94 & 0.01 & 0.98    & 0.98   & 0.95 & 0.96 & 0.95 & 0.95 & 0.01 & 0.99    & 0.99   & 0.96 & 0.97 & 0.94 & 0.95 & 0.06 & 0.99    & 0.99   & 0.95 \\ 
             & \multicolumn{1}{l|}{} & ResNet & 0.95 & 0.96 & 0.96 & 0.10 & 0.97    & 0.93   & 0.92 & 0.96 & 0.94 & 0.95 & 0.00 & 0.96    & 0.96   & 0.94 & 0.97 & 0.90 & 0.93 & 0.01 & 0.94    & 0.95   & 0.93 \\ 
             & \multicolumn{1}{l|}{} & SVM    & 0.95 & 0.97 & 0.96 & 0.00 & 0.97    & 0.93   & 0.95 & 0.76 & 0.72 & 0.74 & 0.00 & 0.91    & 0.83   & 0.75 & 0.84 & 0.91 & 0.87 & 0.05 & 0.95    & 0.92   & 0.87 \\ \hline 
            \multirow{3}{*}{13} & \multicolumn{1}{l|}{\multirow{3}{*}{K9}}  & RF & 0.97 & 0.99 & 0.98 & 0.00 & 1.00    & 1.00   & 0.99 & 0.99 & 1.00 & 0.99 & 0.00 & 1.00    & 1.00   & 0.99 & 0.99 & 1.00 & 1.00 & 0.01 & 1.00    & 1.00   & 0.99 \\ 
             & \multicolumn{1}{l|}{} & ResNet & 0.94 & 1.00 & 0.97 & 0.00 & 0.97    & 0.97   & 0.96 & 0.94 & 0.85 & 0.89 & 0.01 & 0.90    & 0.94   & 0.88 & 0.99 & 1.00 & 0.99 & 0.01 & 0.99    & 1.00   & 0.99 \\ 
             & \multicolumn{1}{l|}{} & SVM    & 0.99 & 1.00 & 0.99 & 0.01 & 0.99    & 0.99   & 0.99 & 0.83 & 0.88 & 0.85 & 0.00 & 0.92    & 0.92   & 0.86 & 0.98 & 0.98 & 0.98 & 0.01 & 0.99    & 0.99   & 0.98 \\ \hline 
            \multirow{3}{*}{14} & \multicolumn{1}{l|}{\multirow{3}{*}{KeePassDroid}} & RF & 0.99 & 1.00 & 1.00 & 0.01 & 1.00    & 1.00   & 1.00 & 1.00 & 0.99 & 1.00 & 0.01 & 1.00    & 1.00   & 0.99 & 0.98 & 0.99 & 0.98 & 0.01 & 1.00    & 1.00   & 0.99 \\ 
             & \multicolumn{1}{l|}{} & ResNet & 0.99 & 0.99 & 0.99 & 0.05 & 0.99    & 0.98   & 0.97 & 0.97 & 0.84 & 0.90 & 0.03 & 0.91    & 0.94   & 0.89 & 0.99 & 1.00 & 0.99 & 0.08 & 1.00    & 0.99   & 0.97 \\ 
             & \multicolumn{1}{l|}{} & SVM    & 0.99 & 0.99 & 0.99 & 0.02 & 0.99    & 0.98   & 0.98 & 0.90 & 0.82 & 0.86 & 0.07 & 0.95    & 0.92   & 0.85 & 0.98 & 1.00 & 0.99 & 0.42 & 1.00    & 0.99   & 0.89 \\ \hline 
            \multirow{3}{*}{15} & \multicolumn{1}{l|}{\multirow{3}{*}{Mms}}     & RF & 0.98 & 0.97 & 0.98 & 0.22 & 1.00    & 0.99   & 0.93 & 0.98 & 0.97 & 0.98 & 0.01 & 0.98    & 0.96   & 0.97 & 0.94 & 1.00 & 0.97 & 0.01 & 0.98    & 0.93   & 0.96 \\ 
             & \multicolumn{1}{l|}{} & ResNet & 0.98 & 0.93 & 0.96 & 0.44 & 0.96    & 0.94   & 0.84 & 0.57 & 0.98 & 0.72 & 0.17 & 0.78    & 0.56   & 0.64 & 0.86 & 0.95 & 0.90 & 0.32 & 0.94    & 0.91   & 0.82 \\ 
             & \multicolumn{1}{l|}{} & SVM    & 0.98 & 0.97 & 0.97 & 0.12 & 0.97    & 0.95   & 0.93 & 0.98 & 0.95 & 0.97 & 0.08 & 0.97    & 0.95   & 0.94 & 0.89 & 0.93 & 0.91 & 0.07 & 0.98    & 0.95   & 0.90 \\ \hline 
            \multirow{3}{*}{16} & \multicolumn{1}{l|}{\multirow{3}{*}{Crosswords}}  & RF & 1.00 & 1.00 & 1.00 & 0.00 & 1.00    & 1.00   & 1.00 & 1.00 & 0.97 & 0.99 & 0.00 & 0.99    & 0.99   & 0.98 & 0.99 & 1.00 & 0.99 & 0.00 & 1.00    & 0.99   & 0.99 \\ 
             & \multicolumn{1}{l|}{} & ResNet & 1.00 & 1.00 & 1.00 & 0.01 & 1.00    & 1.00   & 1.00 & 0.94 & 0.91 & 0.92 & 0.11 & 0.93    & 0.89   & 0.88 & 0.88 & 0.89 & 0.88 & 0.09 & 0.90    & 0.82   & 0.83 \\ 
             & \multicolumn{1}{l|}{} & SVM    & 1.00 & 0.99 & 0.99 & 0.15 & 0.99    & 0.99   & 0.96 & 0.97 & 0.92 & 0.95 & 0.00 & 0.97    & 0.95   & 0.95 & 0.95 & 1.00 & 0.97 & 0.05 & 0.98    & 0.95   & 0.95 \\ \hline
             \multirow{3}{*}{17} & \multicolumn{1}{l|}{\multirow{3}{*}{QuickSearchBox}}  & RF & 0.95 & 0.87 & 0.91 & 0.01 & 0.93 & 0.85 & 0.96 & 0.77 & 0.89 & 0.83 & 0.07 & 0.91 & 0.71 & 0.85 & 0.94 & 0.94 & 0.94 & 0.03 & 0.96 & 0.90 & 1.00\\ 
             & \multicolumn{1}{l|}{} & ResNet & 1.00 &	0.78 &	0.88 &	0.00 &	0.89 &	0.82 & 0.93 & 0.58 &	0.96 &	0.72 &	0.18 &	0.89 &	0.56 & 0.72 & 0.85 & 0.79 &	0.82 &	0.06 &	0.87 &	0.74 & 0.84\\ 
             & \multicolumn{1}{l|}{} & SVM & 0.95 & 0.87 &	0.91 &	0.01 &	0.93 &	0.85 & 0.96 & 0.74 &	0.63 &	0.68 &	0.06 &	0.79 &	0.54 & 0.66 & 0.86 &	0.86 &	0.90 &	0.06 &	0.94 &	0.84 & 0.92\\ \hline
        \end{tabular}
    }
\end{table*}

%% file: Sections/tab/result-table-nocomments.tex
\begin{table*}[hbt!]
    \caption{Singular project vulnerability detection with tokenization without comments and symbols across embeddings and machine learning models.}
    \label{tab:results-no-comms}
    \resizebox{\textwidth}{!}{
\begin{tabular}{|ll|l|cccccc|c|cccccc|c|cccccc|c|}
\cline{4-24}
\multicolumn{1}{c}{} & \multicolumn{1}{c}{} & \multicolumn{1}{c}{} & \multicolumn{7}{|c|}{Bag-of-Words} & \multicolumn{7}{c|}{Word2vec} & \multicolumn{7}{c|}{FastText}\\ \cline{1-24}
\multicolumn{2}{|c|}{Project} & Classifier & P & R & F1 & FPR & ROC AUC & PR AUC & z & P & R & F3 & FPR & ROC AUC & PR AUC & z & P & R & F2 & FPR & ROC AUC & PR AUC & z \\ \cline{1-24}
\multirow{3}{*}{1} & \multicolumn{1}{l|}{\multirow{3}{*}{OWASP}}   & RF & 0.99 & 1.00 & 0.99 & 0.21 & 0.99    & 0.99   & 0.94 & 0.71 & 0.73 & 0.72 & 0.21 & 0.72    & 0.65   & 0.60 & 0.75 & 0.83 & 0.79 & 0.21 & 0.89    & 0.89   & 0.75 \\
 & \multicolumn{1}{l|}{} & Resnet & 0.82 & 1.00 & 0.90 & 0.17 & 0.89    & 0.82   & 0.83 & 0.79 & 0.93 & 0.86 & 0.19 & 0.85    & 0.77   & 0.77 & 0.57 & 0.58 & 0.58 & 0.19 & 0.57    & 0.68   & 0.48 \\
 & \multicolumn{1}{l|}{} & SVM    & 0.99 & 0.99 & 0.99 & 0.24 & 0.99    & 0.99   & 0.93 & 0.88 & 0.90 & 0.89 & 0.31 & 0.89    & 0.84   & 0.78 & 0.60 & 0.79 & 0.68 & 0.19 & 0.68    & 0.67   & 0.58 \\ \cline{1-24}
\multirow{3}{*}{2}  & \multicolumn{1}{l|}{\multirow{3}{*}{Juliet}}  & RF & 0.03 & 0.02 & 0.03 & 0.04 & 0.26    & 0.42   & 0.00 & 0.12 & 0.09 & 0.10 & 0.04 & 0.30    & 0.40   & 0.05 & 0.23 & 0.18 & 0.20 & 0.02 & 0.52    & 0.36   & 0.17 \\
 & \multicolumn{1}{l|}{} & Resnet & 0.33 & 0.05 & 0.08 & 0.04 & 0.35    & 0.42   & 0.11 & 0.22 & 0.09 & 0.13 & 0.03 & 0.43    & 0.40   & 0.12 & 0.47 & 0.34 & 0.37 & 0.07 & 0.54    & 0.54   & 0.34 \\
 & \multicolumn{1}{l|}{} & SVM    & 0.41 & 0.02 & 0.03 & 0.04 & 0.68    & 0.54   & 0.21 & 0.20 & 0.09 & 0.13 & 0.03 & 0.47    & 0.42   & 0.13 & 0.42 & 0.16 & 0.23 & 0.02 & 0.62    & 0.46   & 0.27 \\ \cline{1-24}
\multirow{3}{*}{3} & \multicolumn{1}{l|}{\multirow{3}{*}{Anki-Android}}   & RF & 0.81 & 0.96 & 0.88 & 0.08 & 0.88    & 0.80   & 0.83 & 0.78 & 0.93 & 0.85 & 0.08 & 0.84    & 0.76   & 0.78 & 0.84 & 0.96 & 0.90 & 0.08 & 0.90    & 0.83   & 0.85 \\
 & \multicolumn{1}{l|}{} & Resnet & 0.82 & 1.00 & 0.90 & 0.11 & 0.90    & 0.82   & 0.84 & 0.78 & 0.93 & 0.85 & 0.05 & 0.84    & 0.76   & 0.79 & 0.82 & 0.52 & 0.64 & 0.00 & 0.71    & 0.66   & 0.61 \\
 & \multicolumn{1}{l|}{} & SVM    & 0.81 & 0.96 & 0.88 & 0.16 & 0.90    & 0.82   & 0.82 & 0.80 & 0.89 & 0.84 & 0.13 & 0.84    & 0.76   & 0.77 & 0.77 & 0.85 & 0.81 & 0.09 & 0.65    & 0.57   & 0.66 \\ \cline{1-24}
\multirow{3}{*}{4}  & \multicolumn{1}{l|}{\multirow{3}{*}{Browser}} & RF & 0.94 & 0.91 & 0.93 & 0.04 & 0.94    & 0.89   & 0.90 & 0.94 & 0.94 & 0.94 & 0.04 & 0.95 & 0.90   & 0.91 & 0.93 & 0.84 & 0.89 & 0.04 & 0.96    & 0.87   & 0.87 \\
 & \multicolumn{1}{l|}{} & Resnet & 0.88 & 0.85 & 0.87 & 0.03 & 0.89    & 0.81   & 0.83 & 0.88 & 0.91 & 0.90 & 0.03 & 0.92    & 0.84   & 0.86 & 0.81 & 0.91 & 0.86 & 0.07 & 0.89    & 0.77   & 0.81 \\
 & \multicolumn{1}{l|}{} & SVM    & 0.91 & 0.91 & 0.91 & 0.05 & 0.94    & 0.88   & 0.89 & 0.89 & 1.00 & 0.94 & 0.07 & 0.96    & 0.89   & 0.91 & 0.90 & 0.82 & 0.86 & 0.06 & 0.88    & 0.80   & 0.81 \\ \cline{1-24}
\multirow{3}{*}{5}  & \multicolumn{1}{l|}{\multirow{3}{*}{Calendar}}       & RF & 0.88 & 0.95 & 0.91 & 0.00 & 0.94    & 0.85   & 0.89 & 0.73 & 0.70 & 0.71 & 0.00 & 0.77    & 0.62   & 0.65 & 0.81 & 0.74 & 0.77 & 0.00 & 0.82    & 0.70   & 0.73 \\
 & \multicolumn{1}{l|}{} & Resnet & 0.78 & 0.95 & 0.86 & 0.00 & 0.90    & 0.76   & 0.82 & 0.76 & 0.70 & 0.73 & 0.00 & 0.78    & 0.64   & 0.68 & 0.78 & 0.89 & 0.83 & 0.00 & 0.84    & 0.75   & 0.79 \\
 & \multicolumn{1}{l|}{} & SVM    & 0.88 & 0.95 & 0.91 & 0.00 & 0.94    & 0.85   & 0.89 & 0.71 & 0.74 & 0.72 & 0.00 & 0.78    & 0.62   & 0.67 & 0.77 & 0.74 & 0.76 & 0.00 & 0.80    & 0.67   & 0.71 \\ \cline{1-24}
\multirow{3}{*}{6}  & \multicolumn{1}{l|}{\multirow{3}{*}{Camera}}  & RF & 0.87 & 0.91 & 0.93 & 0.07 & 0.94    & 0.89   & 0.88 & 0.87 & 0.83 & 0.85 & 0.05 & 0.89    & 0.77   & 0.81 & 0.92 & 0.88 & 0.90 & 0.05 & 0.97    & 0.94   & 0.90 \\
 & \multicolumn{1}{l|}{} & Resnet & 0.88 & 0.88 & 0.88 & 0.92 & 0.90    & 0.83   & 0.64 & 0.78 & 0.88 & 0.82 & 0.05 & 0.89    & 0.72   & 0.77 & 0.81 & 0.85 & 0.83 & 0.07 & 0.88    & 0.85   & 0.80 \\
 & \multicolumn{1}{l|}{} & SVM    & 0.91 & 0.91 & 0.91 & 0.04 & 0.94    & 0.88   & 0.89 & 0.88 & 0.92 & 0.90 & 0.07 & 0.93    & 0.93   & 0.88 & 0.81 & 0.81 & 0.81 & 0.08 & 0.89    & 0.74   & 0.76 \\ \cline{1-24}
\multirow{3}{*}{7} & \multicolumn{1}{l|}{\multirow{3}{*}{Connectbot}}     & RF & 1.00 & 1.00 & 1.00 & 0.00 & 1.00    & 1.00   & 1.00 & 1.00 & 1.00 & 1.00 & 0.00 & 1.00    & 1.00   & 1.00 & 1.00 & 1.00 & 1.00 & 0.00 & 1.00    & 1.00   & 1.00 \\
 & \multicolumn{1}{l|}{} & Resnet & 1.00 & 1.00 & 1.00 & 0.00 & 1.00    & 1.00   & 1.00 & 1.00 & 1.00 & 1.00 & 0.00 & 1.00    & 1.00   & 1.00 & 1.00 & 0.82 & 0.90 & 0.00 & 0.91    & 0.93   & 0.90 \\
 & \multicolumn{1}{l|}{} & SVM    & 1.00 & 1.00 & 1.00 & 0.00 & 1.00    & 1.00   & 1.00 & 1.00 & 1.00 & 1.00 & 0.00 & 1.00    & 1.00   & 1.00 & 1.00 & 1.00 & 1.00 & 0.00 & 1.00    & 1.00   & 1.00 \\ \cline{1-24}
\multirow{3}{*}{8}  & \multicolumn{1}{l|}{\multirow{3}{*}{Contacts}}       & RF & 0.85 & 0.96 & 0.90 & 0.06 & 0.98    & 0.95   & 0.90 & 0.93 & 0.91 & 0.92 & 0.05 & 0.94    & 0.88   & 0.89 & 0.96 & 0.89 & 0.92 & 0.06 & 0.93    & 0.89   & 0.89 \\
 & \multicolumn{1}{l|}{} & Resnet & 0.83 & 0.92 & 0.87 & 0.08 & 0.92    & 0.89   & 0.85 & 0.90 & 0.67 & 0.77 & 0.15 & 0.81    & 0.72   & 0.70 & 0.81 & 0.87 & 0.84 & 0.06 & 0.88    & 0.74   & 0.78 \\
 & \multicolumn{1}{l|}{} & SVM    & 0.80 & 0.67 & 0.73 & 0.07 & 0.90    & 0.84   & 0.73 & 0.86 & 0.77 & 0.81 & 0.14 & 0.85    & 0.76   & 0.75 & 0.85 & 0.83 & 0.84 & 0.14 & 0.74    & 0.62   & 0.70 \\ \cline{1-24}
\multirow{3}{*}{9}  & \multicolumn{1}{l|}{\multirow{3}{*}{CoolReader}}     & RF & 1.00 & 0.97 & 0.99 & 0.04 & 0.99    & 0.99   & 0.97 & 1.00 & 1.00 & 1.00 & 0.01 & 1.00    & 1.00   & 1.00 & 1.00 & 0.98 & 0.99 & 0.04 & 0.99    & 0.99   & 0.98 \\
 & \multicolumn{1}{l|}{} & Resnet & 1.00 & 0.97 & 0.99 & 0.04 & 0.99    & 0.99   & 0.97 & 0.98 & 1.00 & 0.99 & 0.01 & 0.99    & 0.98   & 0.98 & 1.00 & 1.00 & 1.00 & 0.10 & 1.00    & 1.00   & 0.98 \\
 & \multicolumn{1}{l|}{} & SVM    & 0.97 & 0.97 & 0.97 & 0.09 & 0.96    & 0.93   & 0.94 & 0.98 & 1.00 & 0.99 & 0.00 & 0.99    & 0.98   & 0.98 & 1.00 & 0.98 & 0.99 & 0.03 & 0.99    & 0.99   & 0.98 \\ \cline{1-24}
\multirow{3}{*}{10} & \multicolumn{1}{l|}{\multirow{3}{*}{DeskClock}}      & RF & 0.89 & 1.00 & 0.94 & 0.02 & 0.97    & 0.89   & 0.92 & 0.91 & 0.83 & 0.87 & 0.02 & 0.88    & 0.83   & 0.84 & 0.92 & 0.79 & 0.85 & 0.02 & 0.93    & 0.87   & 0.84 \\
 & \multicolumn{1}{l|}{} & Resnet & 0.98 & 1.00 & 0.89 & 0.01 & 0.94    & 0.80   & 0.91 & 0.75 & 0.75 & 0.75 & 0.01 & 0.77    & 0.68   & 0.69 & 0.50 & 0.07 & 0.13 & 0.01 & 0.49    & 0.54   & 0.23 \\
 & \multicolumn{1}{l|}{} & SVM    & 0.89 & 1.00 & 0.94 & 0.01 & 0.97    & 0.89   & 0.93 & 0.83 & 0.83 & 0.83 & 0.02 & 0.85    & 0.77   & 0.79 & 0.80 & 0.57 & 0.67 & 0.02 & 0.86    & 0.88   & 0.71 \\ \cline{1-24}
\multirow{3}{*}{11} & \multicolumn{1}{l|}{\multirow{3}{*}{Email}}   & RF & 0.97 & 0.98 & 0.97 & 0.50 & 1.00    & 1.00   & 0.86 & 0.93 & 0.99 & 0.96 & 0.00 & 0.98    & 0.97   & 0.96 & 0.97 & 0.98 & 0.97 & 0.00 & 0.97    & 0.96   & 0.97 \\
 & \multicolumn{1}{l|}{} & Resnet & 0.93 & 1.00 & 0.96 & 0.41 & 0.95    & 0.97   & 0.86 & 0.97 & 0.75 & 0.85 & 0.47 & 0.86    & 0.87   & 0.72 & 0.91 & 0.99 & 0.95 & 0.45 & 0.93    & 0.91   & 0.82 \\
 & \multicolumn{1}{l|}{} & SVM    & 0.96 & 0.85 & 0.90 & 0.50 & 0.96    & 0.97   & 0.80 & 0.97 & 0.98 & 0.97 & 0.45 & 0.97    & 0.96   & 0.86 & 0.94 & 0.95 & 0.94 & 0.45 & 0.93    & 0.92   & 0.82 \\ \cline{1-24}
\multirow{3}{*}{12} & \multicolumn{1}{l|}{\multirow{3}{*}{FBReaderJ}}      & RF & 0.96 & 0.96 & 0.97 & 0.01 & 0.98    & 0.95   & 0.96 & 0.98 & 0.95 & 0.97 & 0.02 & 0.97    & 0.95   & 0.95 & 0.97 & 0.95 & 0.96 & 0.02 & 1.00    & 0.99   & 0.96 \\
 & \multicolumn{1}{l|}{} & Resnet & 0.96 & 0.96 & 0.96 & 0.01 & 0.97    & 0.93   & 0.95 & 0.95 & 0.90 & 0.93 & 0.00 & 0.94    & 0.89   & 0.91 & 0.92 & 0.82 & 0.87 & 0.01 & 0.90    & 0.90   & 0.86 \\
 & \multicolumn{1}{l|}{} & SVM    & 0.95 & 0.97 & 0.96 & 0.02 & 0.97    & 0.93   & 0.94 & 0.93 & 0.85 & 0.89 & 0.01 & 0.91    & 0.83   & 0.86 & 0.75 & 0.50 & 0.60 & 0.01 & 0.86    & 0.69   & 0.62 \\ \cline{1-24}
\multirow{3}{*}{13} & \multicolumn{1}{l|}{\multirow{3}{*}{K9Mail}}  & RF & 0.99 & 1.00 & 0.99 & 0.00 & 0.99    & 0.99   & 0.99 & 0.98 & 0.99 & 0.98 & 0.00 & 1.00    & 1.00   & 0.99 & 0.99 & 0.99 & 0.99 & 0.00 & 1.00    & 1.00   & 0.99 \\
 & \multicolumn{1}{l|}{} & Resnet & 0.99 & 0.99 & 0.99 & 0.00 & 0.99    & 0.99   & 0.99 & 1.00 & 0.94 & 0.97 & 0.00 & 0.97    & 0.97   & 0.96 & 0.92 & 0.91 & 0.91 & 0.01 & 0.91    & 0.94   & 0.90 \\
 & \multicolumn{1}{l|}{} & SVM    & 0.99 & 1.00 & 1.00 & 0.01 & 0.99    & 0.99   & 0.99 & 0.98 & 1.00 & 0.99 & 0.00 & 0.98    & 0.97   & 0.98 & 0.77 & 0.88 & 0.82 & 0.00 & 0.85    & 0.80   & 0.79 \\ \cline{1-24}
\multirow{3}{*}{14} & \multicolumn{1}{l|}{\multirow{3}{*}{KeePassAndroid}} & RF & 1.00 & 1.00 & 1.00 & 0.01 & 1.00    & 1.00   & 1.00 & 0.99 & 1.00 & 1.00 & 0.01 & 1.00    & 0.99   & 0.99 & 0.99 & 1.00 & 1.00 & 0.01 & 1.00    & 1.00   & 1.00 \\
 & \multicolumn{1}{l|}{} & Resnet & 1.00 & 1.00 & 1.00 & 0.03 & 1.00    & 1.00   & 0.99 & 0.99 & 1.00 & 1.00 & 0.03 & 1.00    & 0.99   & 0.99 & 1.00 & 0.99 & 0.99 & 0.03 & 0.99    & 1.00   & 0.98 \\
 & \multicolumn{1}{l|}{} & SVM    & 0.98 & 0.97 & 0.97 & 0.03 & 1.00    & 1.00   & 0.97 & 0.99 & 1.00 & 1.00 & 0.03 & 1.00    & 0.99   & 0.99 & 0.83 & 0.86 & 0.84 & 0.01 & 0.92    & 0.84   & 0.83 \\ \cline{1-24}
\multirow{3}{*}{15} & \multicolumn{1}{l|}{\multirow{3}{*}{MMS}}     & RF & 0.98 & 0.97 & 0.97 & 0.01 & 0.98    & 0.96   & 0.96 & 0.96 & 0.98 & 0.97 & 0.01 & 0.98    & 0.95   & 0.96 & 0.96 & 0.98 & 0.97 & 0.01 & 0.98    & 0.95   & 0.96 \\
 & \multicolumn{1}{l|}{} & Resnet & 0.98 & 0.91 & 0.94 & 0.28 & 0.95    & 0.96   & 0.87 & 0.96 & 0.67 & 0.79 & 0.00 & 0.82    & 0.76   & 0.76 & 0.91 & 0.95 & 0.93 & 0.00 & 0.95    & 0.89   & 0.92 \\
 & \multicolumn{1}{l|}{} & SVM    & 0.98 & 0.97 & 0.97 & 0.12 & 0.98    & 0.96   & 0.94 & 0.96 & 0.98 & 0.97 & 0.04 & 0.97    & 0.94   & 0.95 & 0.96 & 0.98 & 0.97 & 0.09 & 0.98    & 0.95   & 0.94 \\ \cline{1-24}
\multirow{3}{*}{16} & \multicolumn{1}{l|}{\multirow{3}{*}{Xwords}}  & RF & 1.00 & 1.00 & 1.00 & 0.00 & 1.00    & 1.00   & 1.00 & 0.99 & 1.00 & 0.99 & 0.00 & 0.99    & 0.99   & 0.99 & 1.00 & 1.00 & 1.00 & 0.00 & 1.00    & 1.00   & 1.00 \\
 & \multicolumn{1}{l|}{} & Resnet & 1.00 & 0.99 & 0.99 & 0.00 & 0.99    & 0.99   & 0.99 & 0.86 & 1.00 & 0.92 & 0.00 & 0.92    & 0.86   & 0.90 & 0.95 & 0.25 & 0.40 & 0.01 & 0.62    & 0.62   & 0.49 \\
 & \multicolumn{1}{l|}{} & SVM    & 1.00 & 0.99 & 0.99 & 0.01 & 0.99    & 0.99   & 0.99 & 1.00 & 0.99 & 0.99 & 0.00 & 0.99    & 0.98   & 0.99 & 1.00 & 1.00 & 1.00 & 0.00 & 1.00    & 1.00   & 1.00 \\ \cline{1-24}
 \multirow{3}{*}{17} & \multicolumn{1}{l|}{\multirow{3}{*}{QuickSearchBox}}  & RF & 0.95 & 0.87 &	0.91 &	0.01 &	0.93 &	0.85 & 0.96 &	0.88 &	0.81 &	0.85 &	0.03 &	0.89 &	0.76 & 0.87 &	0.93 &	0.95 &	0.94 &	0.03 &	0.96 &	0.90 & 1.00\\ 
 & \multicolumn{1}{l|}{} & Resnet  & 0.95 &	0.91 &	0.93 &	0.01 &	0.95 &	0.89 & 0.99 &	0.77 &	0.89 &	0.83 &	0.07 &	0.91 &	0.71 & 0.85 &	0.86 &	0.99 &	0.92 &	0.07 &	0.96 &	0.86 & 0.97 \\
  & \multicolumn{1}{l|}{} & SVM  & 0.95 &	0.87 &	0.91 &	0.01 &	0.93 & 0.85 & 0.96 &	0.89 &	0.89 &	0.89 &	0.03 &	0.90 &	0.82 & 0.92 &	0.85 &	0.88 &	0.86 &	0.07 &	0.89 &	0.77 & 0.88\\\hline
\end{tabular}
}
\end{table*}

%% file: Sections/tab/results-table-arch.tex
\begin{table*}[hbt!]
\caption{Singular project vulnerability detection with Bag-of-word and the Architectural metrics}
\label{tab:results-arch}
    \resizebox{\textwidth}{!}{
\begin{tabular}{|ll|l|cccccc|c|cccccc|c|}
\cline{4-17}
\multicolumn{3}{c}{}& \multicolumn{7}{|c|}{Metrics only} & \multicolumn{7}{c|}{Metrics + bag-of-words} \\ \hline
\multicolumn{2}{|c|}{Project} & Classifier & P & R & F1 & FPR & ROC AUC & PR AUC & z & P & R & F1 & FPR & ROC AUC & PR AUC & z \\ \hline
\multirow{3}{*}{1} & \multicolumn{1}{l|}{\multirow{3}{*}{OWASP}} &  RF & 0.66 & 0.78 & 0.71 & 0.38 & 0.70 & 0.62 & 0.56 & 0.82 & 0.85 & 0.84 & 0.17 & 0.95 & 0.96 & 0.73 \\  
& \multicolumn{1}{l|}{} & Resnet & 0.48 & 1.00 & 0.65 & 1.00 & 0.50 & 0.48 & 0.32 & 0.70 & 0.89 & 0.79 & 0.34 & 0.77 & 0.82 & 0.51 \\  
& \multicolumn{1}{l|}{} & SVM & 0.57 & 0.93 & 0.70 & 0.66 & 0.64 & 0.56 & 0.48 & 0.67 & 0.74 & 0.70 & 0.74 & 0.82 & 0.85 & 0.30 \\ \hline
\multirow{3}{*}{2} & \multicolumn{1}{l|}{\multirow{3}{*}{Juliet}} &  RF & 0.50 & 0.41 & 0.45 & 0.23 & 0.59 & 0.42 & 0.33 & 1.00 & 0.88 & 0.93 & 0.00 & 0.94 & 0.92 & 0.88 \\  
& \multicolumn{1}{l|}{} & Resnet & 0.35 & 0.97 & 0.52 & 1.00 & 0.48 & 0.35 & 0.21 & 1.00 & 0.81 & 0.90 & 0.00 & 0.91 & 0.88 & 0.82 \\  
& \multicolumn{1}{l|}{} & SVM & 0.00 & 0.00 & 0.00 & 0.00 & 0.50 & 0.36 & 0.01 & 1.00 & 0.84 & 0.92 & 0.00 & 0.94 & 0.92 & 0.86 \\ \hline
\multirow{3}{*}{3} & \multicolumn{1}{l|}{\multirow{3}{*}{Anki-Android}} &  RF & 0.62 & 0.73 & 0.67 & 0.26 & 0.74 & 0.55 & 0.55 & 0.87 & 0.91 & 0.89 & 0.08 & 0.92 & 0.82 & 0.76 \\  
& \multicolumn{1}{l|}{} & Resnet & 0.36 & 1.00 & 0.53 & 1.00 & 0.50 & 0.36 & 0.23 & 0.83 & 0.86 & 0.84 & 0.10 & 0.88 & 0.76 & 0.67 \\  
& \multicolumn{1}{l|}{} & SVM & 0.71 & 0.23 & 0.34 & 0.05 & 0.50 & 0.36 & 0.32 & 0.88 & 0.95 & 0.91 & 0.08 & 0.94 & 0.85 & 0.81 \\ \hline
\multirow{3}{*}{4} & \multicolumn{1}{l|}{\multirow{3}{*}{Browser}} &  RF & 0.72 & 0.62 & 0.67 & 0.16 & 0.73 & 0.60 & 0.59 & 0.94 & 0.91 & 0.93 & 0.04 & 0.94 & 0.89 & 0.84 \\  
& \multicolumn{1}{l|}{} & Resnet & 0.00 & 0.00 & 0.00 & 0.00 & 0.50 & 0.40 & 0.02 & 0.91 & 0.91 & 0.91 & 0.06 & 0.93 & 0.87 & 0.81 \\  
& \multicolumn{1}{l|}{} & SVM & 0.00 & 0.00 & 0.00 & 0.00 & 0.50 & 0.40 & 0.02 & 0.89 & 0.94 & 0.93 & 0.06 & 0.94 & 0.88 & 0.83 \\ \hline
\multirow{3}{*}{5} & \multicolumn{1}{l|}{\multirow{3}{*}{Calendar}} &  RF & 1.00 & 0.94 & 0.97 & 0.00 & 0.97 & 0.98 & 1.00 & 1.00 & 1.00 & 1.00 & 0.00 & 1.00 & 1.00 & 1.00 \\  
& \multicolumn{1}{l|}{} & Resnet & 0.90 & 0.53 & 0.67 & 0.08 & 0.72 & 0.75 & 0.66 & 0.75 & 0.88 & 0.81 & 0.42 & 0.73 & 0.85 & 0.50 \\  
& \multicolumn{1}{l|}{} & SVM & 0.90 & 0.53 & 0.67 & 0.08 & 0.72 & 0.75 & 0.66 & 1.00 & 0.88 & 0.94 & 0.00 & 1.00 & 1.00 & 0.94 \\ \hline
\multirow{3}{*}{6} & \multicolumn{1}{l|}{\multirow{3}{*}{Camera}} &  RF & 0.57 & 0.60 & 0.59 & 0.20 & 0.70 & 0.46 & 0.47 & 0.90 & 0.96 & 0.93 & 0.05 & 0.96 & 0.88 & 0.85 \\  
& \multicolumn{1}{l|}{} & Resnet & 0.39 & 0.56 & 0.46 & 0.39 & 0.59 & 0.35 & 0.28 & 0.84 & 0.88 & 0.86 & 0.07 & 0.90 & 0.77 & 0.71 \\  
& \multicolumn{1}{l|}{} & SVM & 0.00 & 0.00 & 0.00 & 0.00 & 0.50 & 0.30 & 0.00 & 0.90 & 0.96 & 0.93 & 0.05 & 0.96 & 0.88 & 0.85 \\ \hline
\multirow{3}{*}{7} & \multicolumn{1}{l|}{\multirow{3}{*}{Connectbot}} &  RF & 0.80 & 0.76 & 0.78 & 0.15 & 0.80 & 0.71 & 0.71 & 1.00 & 1.00 & 1.00 & 0.00 & 1.00 & 1.00 & 1.00 \\  
& \multicolumn{1}{l|}{} & Resnet & 0.45 & 0.46 & 0.45 & 0.46 & 0.50 & 0.45 & 0.26 & 1.00 & 0.97 & 0.99 & 0.00 & 0.99 & 0.99 & 0.98 \\  
& \multicolumn{1}{l|}{} & SVM & 0.65 & 0.54 & 0.61 & 0.47 & 0.43 & 0.04 & 0.25 & 0.92 & 0.93 & 0.93 & 0.06 & 0.98 & 0.97 & 0.88 \\ \hline
\multirow{3}{*}{8} & \multicolumn{1}{l|}{\multirow{3}{*}{Contacts}} &  RF & 0.89 & 1.00 & 0.94 & 0.06 & 0.97 & 0.89 & 0.95 & 0.89 & 1.00 & 0.94 & 0.06 & 0.89 & 0.92 & 0.85 \\  
& \multicolumn{1}{l|}{} & Resnet & 0.31 & 1.00 & 0.47 & 1.00 & 0.50 & 0.31 & 0.19 & 0.80 & 1.00 & 0.89 & 0.11 & 0.94 & 0.80 & 0.76 \\  
& \multicolumn{1}{l|}{} & SVM & 0.54 & 0.88 & 0.67 & 0.33 & 0.77 & 0.51 & 0.55 & 0.89 & 1.00 & 0.94 & 0.06 & 0.89 & 0.92 & 0.85 \\ \hline
\multirow{3}{*}{9} & \multicolumn{1}{l|}{\multirow{3}{*}{CoolReader}} &  RF & 0.83 & 0.86 & 0.84 & 0.20 & 0.83 & 0.79 & 0.77 & 0.99 & 0.96 & 0.97 & 0.01 & 0.97 & 0.97 & 0.94 \\  
& \multicolumn{1}{l|}{} & Resnet & 0.61 & 0.84 & 0.71 & 0.62 & 0.61 & 0.60 & 0.48 & 0.98 & 0.96 & 0.97 & 0.03 & 0.97 & 0.96 & 0.93 \\  
& \multicolumn{1}{l|}{} & SVM & 0.63 & 0.77 & 0.69 & 0.52 & 0.62 & 0.61 & 0.49 & 0.98 & 0.96 & 0.97 & 0.03 & 0.97 & 0.96 & 0.93 \\ \hline
\multirow{3}{*}{10} & \multicolumn{1}{l|}{\multirow{3}{*}{DeskClock}} &  RF & 0.83 & 0.68 & 0.75 & 0.06 & 0.81 & 0.67 & 0.71 & 0.98 & 0.96 & 0.97 & 0.01 & 0.97 & 0.95 & 0.94 \\  
& \multicolumn{1}{l|}{} & Resnet & 0.46 & 0.53 & 0.49 & 0.29 & 0.62 & 0.39 & 0.35 & 0.98 & 0.91 & 0.94 & 0.01 & 0.95 & 0.92 & 0.89 \\  
& \multicolumn{1}{l|}{} & SVM & 0.00 & 0.00 & 0.00 & 0.00 & 0.50 & 0.32 & 0.00 & 0.97 & 0.97 & 0.97 & 0.01 & 0.97 & 0.94 & 0.93 \\ \hline
\multirow{3}{*}{11} & \multicolumn{1}{l|}{\multirow{3}{*}{Email}} &  RF & 0.00 & 0.00 & 0.00 & 0.00 & 0.50 & 0.42 & 0.03 & 1.00 & 1.00 & 1.00 & 0.00 & 1.00 & 1.00 & 1.00 \\  
& \multicolumn{1}{l|}{} & Resnet & 0.41 & 1.00 & 0.60 & 1.00 & 0.50 & 0.42 & 0.28 & 1.00 & 0.98 & 0.99 & 0.00 & 0.99 & 0.99 & 0.98 \\  
& \multicolumn{1}{l|}{} & SVM & 0.00 & 0.00 & 0.00 & 0.00 & 0.50 & 0.42 & 0.03 & 1.00 & 1.00 & 1.00 & 0.00 & 1.00 & 1.00 & 1.00 \\ \hline
\multirow{3}{*}{12} & \multicolumn{1}{l|}{\multirow{3}{*}{FBReaderJ}} &  RF & 0.80 & 0.82 & 0.81 & 0.20 & 0.81 & 0.74 & 0.73 & 1.00 & 1.00 & 1.00 & 0.00 & 1.00 & 1.00 & 1.00 \\  
& \multicolumn{1}{l|}{} & Resnet & 0.47 & 0.38 & 0.42 & 0.41 & 0.48 & 0.48 & 0.25 & 1.00 & 0.98 & 0.99 & 0.00 & 0.99 & 0.99 & 0.98 \\  
& \multicolumn{1}{l|}{} & SVM & 0.64 & 0.86 & 0.74 & 0.47 & 0.70 & 0.62 & 0.56 & 0.99 & 1.00 & 0.99 & 0.01 & 0.99 & 0.99 & 0.98 \\ \hline
\multirow{3}{*}{13} & \multicolumn{1}{l|}{\multirow{3}{*}{K9Mail}} &  RF & 0.90 & 0.83 & 0.86 & 0.07 & 0.88 & 0.82 & 0.84 & 0.99 & 0.99 & 0.99 & 0.01 & 0.99 & 0.98 & 0.98 \\  
& \multicolumn{1}{l|}{} & Resnet & 0.71 & 0.27 & 0.39 & 0.08 & 0.59 & 0.51 & 0.39 & 0.46 & 1.00 & 0.63 & 0.90 & 0.55 & 0.46 & 0.00 \\  
& \multicolumn{1}{l|}{} & SVM & 0.00 & 0.00 & 0.00 & 0.00 & 0.50 & 0.43 & 0.03 & 0.99 & 0.99 & 0.99 & 0.01 & 0.99 & 0.98 & 0.98 \\ \hline
\multirow{3}{*}{14} & \multicolumn{1}{l|}{\multirow{3}{*}{KeePassAndroid}} &  RF & 0.86 & 0.83 & 0.84 & 0.07 & 0.88 & 0.77 & 0.81 & 0.98 & 0.97 & 0.97 & 0.01 & 0.98 & 0.96 & 0.95 \\  
& \multicolumn{1}{l|}{} & Resnet & 0.43 & 0.71 & 0.54 & 0.45 & 0.63 & 0.40 & 0.36 & 0.95 & 0.95 & 0.96 & 0.01 & 0.97 & 0.95 & 0.92 \\  
& \multicolumn{1}{l|}{} & SVM & 0.57 & 0.55 & 0.56 & 0.20 & 0.68 & 0.68 & 0.50 & 0.98 & 0.97 & 0.97 & 0.01 & 0.97 & 0.95 & 0.94 \\ \hline
\multirow{3}{*}{15} & \multicolumn{1}{l|}{\multirow{3}{*}{MMS}} &  RF & 0.52 & 0.89 & 0.65 & 0.82 & 0.54 & 0.51 & 0.37 & 1.00 & 1.00 & 1.00 & 0.00 & 1.00 & 1.00 & 1.00 \\  
& \multicolumn{1}{l|}{} & Resnet & 0.51 & 0.49 & 0.50 & 0.46 & 0.51 & 0.50 & 0.31 & 1.00 & 1.00 & 1.00 & 0.00 & 1.00 & 1.00 & 1.00 \\  
& \multicolumn{1}{l|}{} & SVM & 0.50 & 1.00 & 0.66 & 1.00 & 0.50 & 0.50 & 0.33 & 0.99 & 0.99 & 0.99 & 0.01 & 0.99 & 0.99 & 0.98 \\ \hline
\multirow{3}{*}{16} & \multicolumn{1}{l|}{\multirow{3}{*}{Xwords}} &  RF & 0.86 & 0.87 & 0.87 & 0.16 & 0.86 & 0.82 & 0.82 & 1.00 & 1.00 & 1.00 & 0.00 & 1.00 & 1.00 & 1.00 \\  
& \multicolumn{1}{l|}{} & Resnet & 0.61 & 0.92 & 0.74 & 0.65 & 0.64 & 0.61 & 0.51 & 0.98 & 0.95 & 0.96 & 0.03 & 0.96 & 0.98 & 0.93 \\  
& \multicolumn{1}{l|}{} & SVM & 0.75 & 0.67 & 0.70 & 0.25 & 0.71 & 0.68 & 0.60 & 1.00 & 0.93 & 0.96 & 0.00 & 0.99 & 0.99 & 0.95 \\ \hline

 \multirow{3}{*}{17} & \multicolumn{1}{l|}{\multirow{3}{*}{QuickSearchBox}}  & RF & 0.65 &	0.74 &	0.69 &	0.08 &	0.83 &	0.53 & 0.67 & 0.95 & 0.87 &	0.91 &	0.01 & 0.93 &	0.85 & 0.96\\ 
 & \multicolumn{1}{l|}{} & Resnet & 0.50 &	0.04 &	0.08 &	0.01 &	0.52 &	0.19 & 0.16 & 0.95 &	0.87 &	0.91 &	0.01 &	0.93 &	0.85 & 0.96  \\ 
 & \multicolumn{1}{l|}{} & SVM & 0.00 &	0.00 &	0.00 &	0.00 &	0.50 &	0.18 & 0.00 &	0.95 &	0.87 &	0.91 &	0.01 &	0.93 &	0.85 & 0.96 \\ \hline
\end{tabular}
}
\end{table*}

%% file: Sections/Threats.tex
We summarize  aspects of threats to validity, including (1) dataset and limited architectural metrics relative to internal validity; and (3) domains of experiments relative to external validity. 

\textbf{Dataset.} We choose to use publicly available datasets that were previously labelled. The OWASP dataset and the Juliet dataset contain both source code files and vulnerability labels. The Android Study dataset only includes information on the tagged file but without the source code files.  We retrieved the source code according to the file names and project versions. We only used Java source code because many C++ source codes lack data labels. The number of projects examined for transferability is still limited to reach a statistically significant conclusion. 

The vulnerable code labels for the Android Study project followed the data in ~\cite{6860243}. The labels are determined by Fortify~\cite{fortify}. It has been recognized that static code analysis tools may contain false positive labels. In the literature \cite{Perl:2015:VFP:2810103.2813604, Zhou:2017:AIS:3106237.3117771} path and commit data have been mined to identify vulnerable code.    In the security development and operation process, this was addressed by manual correction. In this work, we focus on the factors that contribute to the baseline, and thus assume that the labels are of stable quality.

\textbf{Architectural Metrics.} The token-based feature representation is considered a flattened structure. Such a token-based feature representation is combined with aggregated architectural metrics. The architecture metrics have not contributed significantly to the learning, which indicates either the current learning representation has not utilized the architectural metrics in the optimal embedding or other kinds of learning models should be applied to architectural metrics. This remains further research.

\begin{table}[h!]
    \centering
    \caption{Cross domain comparison to observe how transferable the vulnerability signature is}
    \label{tab:cross-domain}
    \resizebox{\columnwidth}{!}{
    \begin{tabular}{|l||c|c|c|}
        \hline
        \backslashbox{Model}{Predict}& Juliet & OWASP &  Android \\ \hline\hline
        Juliet & Table \ref{tab:results-all-tokens} & \begin{tabular}[c]{@{}l@{}}P: 0.54 \\ R: 0.77\end{tabular} &  \begin{tabular}[c]{@{}l@{}}P: 0.44 \\ R: 0.53  \end{tabular} \\ \hline
        OWASP & \begin{tabular}[c]{@{}l@{}}P: 0.4 \\ R: 0.8\end{tabular} & Table \ref{tab:results-all-tokens} & \multicolumn{1}{p{22mm}|}{1 out of 15 project (precision and recall greater than 0.7) P: 0.74 \newline R: 0.39 }\\ \hline
        Android & \begin{tabular}[c]{@{}l@{}}P: 0.4 \\ R: 0.46\end{tabular} & \begin{tabular}[c]{@{}l@{}}P: 0.49 \\ R: 0.64\end{tabular} & Table \ref{tab:results-all-tokens} \\ \hline
        \end{tabular}
        }
\end{table}

\textbf{Cross Domain Validation.} The cross-domain validation means training a model on datasets from one domain and predicting vulnerabilities on datasets from another domain. The three datasets presented in this paper---OWASP, Juliet, and Android---are from different domains. The previous discussion of the vulnerable files and types in Table~\ref{tab:owasp-vul}, Table~\ref{tab:juliet-vul} and Table~\ref{tab:projects-datasets} show this heterogeneity. Table~\ref{tab:cross-domain} shows the learning performance has degraded. A key contributing factor is the disparateness of vulnerability signatures.
Our cross domain validation is also limited because we could only assess three different domains.

%% file: Sections/Conclusion.tex
\section{CONCLUSION}
This paper proposes to reveal the most contributing factors for detecting software vulnerabilities. The observations from 17 Java projects and over 400 experiments lead to a baseline model on how to choose tokenization techniques, embedding methods, and machine learning models. The baseline model with under cross-validation training approach on the same project domain achieves comparable and slightly better learning performance to the models using deep learning networks. This provides the reference as the least learning performance that a future vulnerability detection approach should achieve. We observe that cross-domain learning is subject to the extent of vulnerability signature disparateness. We envision a promising research direction that integrates transfer learning techniques to a software DevOps process and feeds target domain inputs to augment the training from the source domain.


\section{ACKNOWLEDGMENTS}
We acknowledge colleague Jincheng Sun for providing the ResNet model.